\documentclass[
    aip,
    jcp,
    reprint,
    amssymb,
    twocolumn,
    preprintnumbers,
    eprint=false,
    doi=false
]{revtex4-1}

\usepackage{graphicx}
\usepackage{dcolumn}
\usepackage[utf8]{inputenc}
\DeclareUnicodeCharacter{2010}{-}
\usepackage[breaklinks,hidelinks,pdfusetitle]{hyperref}
\usepackage{amsmath}
\usepackage{siunitx}
\DeclareSIUnit{\calorie}{cal}

\newcommand{\ignore}[1]{}

\newcommand{\precite}{\,}
\newcommand{\prefig}{Fig.\ }
\newcommand{\prefigs}{Figs.\ }
\newcommand{\Prefig}{Figure\ }

\newcommand{\presec}{Sec.\,}
\newcommand{\Presec}{Section\,}
\newcommand{\preapp}{App.\,}
\newcommand{\pretab}{Tab.\,}
\newcommand{\preonlineref}{Ref.\,}

\newcommand{\sta}{{\mathcal{A}}}
\newcommand{\stb}{{\mathcal{B}}}
\newcommand{\sts}{{\mathcal{S}}}

\newcommand{\defect}[1]{{\sffamily #1}}

\newcommand{\figleft}{Left:~}
\newcommand{\figright}{Right:~}

\newcommand{\ta}{{t_\mathcal{A}}}
\newcommand{\tm}{{t_\text{m}}}
\newcommand{\te}{{t_\text{E}}}

\renewcommand{\vec}[1]{\mathbf{#1}}

\begin{document}

\preprint{DRAFT-\today}

\title{The microscopic mechanism of bulk melting of ice}


\author{Clemens Moritz}
\affiliation{Faculty of Physics, University of Vienna, Boltzmanngasse 5, 1090 Vienna, Austria}
\author{Phillip L. Geissler}
\affiliation{Department of Chemistry, University of California, Berkeley, California 94720}
\author{Christoph Dellago}
\email[]{christoph.dellago@univie.ac.at}
\affiliation{Faculty of Physics, University of Vienna, Boltzmanngasse 5, 1090 Vienna, Austria}
\affiliation{Erwin Schrödinger Institute for Mathematics and Physics, Boltzmanngasse 9, 1090, Vienna, Austria}

\date{\today}

\begin{abstract}
We study the initial stages of homogeneous melting of a hexagonal ice crystal at coexistence and at moderate superheating. Our trajectory-based computer simulation approach provides a comprehensive picture of the events that lead to melting; from the initial accumulation of 5+7 defects, via the formation of L-D and interstitial-vacancy pairs, to the formation of a liquid nucleus. Of the different types of defects that we observe to be involved in melting, a particular kind of 5+7 type defect (type 5) plays a prominent role as it often forms prior to the formation of the initial liquid nucleus and close to the site where the nucleus forms. Hence, like other solids, ice homogeneously melts via the prior accumulation of defects.
\end{abstract}


\maketitle

\section{Introduction}\label{sec:intro}
In this paper we present a computer simulation study of the initial stages of melting of hexagonal water ice (Ice Ih) at ambient pressure and superheating up to $11\%$ above the melting point in a regime where the formation of a liquid nucleus of sufficient size is the rate limiting step. Our analysis of ensembles of trajectories yields a detailed, time-resolved picture of the different dynamical pathways that lead to the formation of such a liquid nucleus and the role that different types of defects play. We find that prior to melting a number of so-called \defect{5+7} defects and larger defect structures in the hydrogen-bond network accumulate in the volume that later becomes the liquid nucleus and that the size and number of these defects becomes larger as the degree of superheating is reduced.

The microscopic mechanisms that lead to melting of crystalline solids are a longstanding subject of solid state theory. In most situations melting originates at the surfaces of a crystal\precite\cite{Cahn1986}, however, under particular circumstances a solid may melt from within the bulk of a crystal as opposed to its surfaces\precite\cite{Mei2007} (so called homogeneous melting). For example, micrometer sized single crystal spheres made of silver melt homogeneously if their surfaces are covered by layers of gold atoms. This shell of atoms suppresses surface melting\precite\cite{Daeges1986} by having a higher melting point while forming a lattice that is compatible with the silver lattice. Using this method, substantial superheating of the Ag lattice can be achieved that is preempted by surface melting in a conventional setting.

A multitude of theories exist that put increasingly stringent limits on the amount of superheating a crystal can be subjected to before it becomes mechanically unstable\precite\cite{Lindemann1910,Born1939,Mori1974,Fecht1988,Edgeworth1989,Jin2001c,Belonoshko2006}. However, in equilibrium all of these instabilities are preempted by thermal melting of the crystal via a nucleation and growth mechanism\precite\cite{Lu1998}. In many recent studies preexisting defects have been found to play a major role\precite\cite{Fecht1992,Lu1998,Forsblom2005,Donadio2005,Luo2007,Bai2008,Wang2012a,Mochizuki2013,Samanta2014a,Liang2014} in melting mechanisms and to significantly influence the stability limits of crystals.
\begin{figure}[tbp]
  \begin{center}
    \includegraphics[width=1.0\columnwidth]{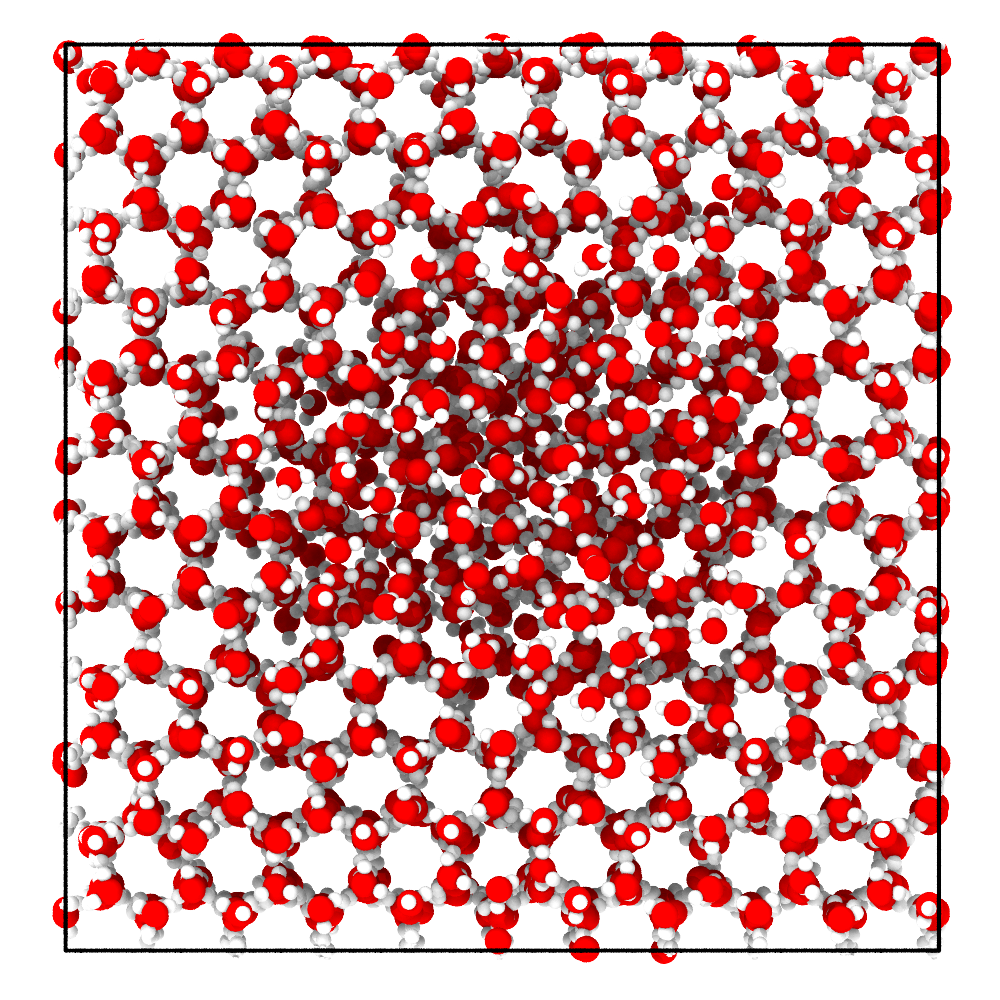}
  \end{center}
  \caption{Example configuration containing a spherical liquid cluster. For clarity, the potential energy of the configuration has been locally minimized. The configuration has been prepared by selecting the molecules in a sphere of radius $\SI{1.5}{\nano\meter}$ and melting them by heating. Afterwards the configuration is briefly equilibrated at a temperature of $\SI{303}{\kelvin}$.}\label{fig:example_liquid_cluster}
\end{figure}

Melting of ice Ih---the ordinary form of ice that can be seen around the liquid nucleus in \prefig\ref{fig:example_liquid_cluster}---is a particularly interesting example of a melting crystal, as its open structure is held together by a network of hydrogen bonds. At temperatures around the ambient pressure melting point, these bonds can rupture and form with relative ease compared to, e.g., covalent bonds, and, hence, a multitude of hydrogen bonding defects occurs in ice under these conditions. While ice usually melts heterogeneously, homogeneous nucleation has been induced by internal heating. This is achieved by exciting the OH stretching mode of water with IR laser light\precite\cite{Iglev2006,Schmeisser2006,Schmeisser2007,Schmeisser2007b,Fanetti2019}. Such experiments strongly suggest that defects play a key role in determining the stability of ice crystals\precite\cite{Schmeisser2007}.

Previous computer simulation studies of ice melting support the finding that defects play an important role.
\textcite{Donadio2005} investigated the free energy landscape of melting at the melting point, finding that so-called \defect{5+7} defects form a minimum in the free energy landscape (see \presec\ref{sec:known_knowns} for a description of the various defect types found in ice). Furthermore, they observed the formation of large defect structures in the hydrogen bond network that involve on the order of 50 molecules. \textcite{Mochizuki2013} studied spontaneous melting under higher superheating (around $19\%$ above melting temperature) where melting events occur spontaneously in simulations within a timeframe of a few nanoseconds. They identified the formation of separated defect pairs (either interstitial-vacancy or \defect{L-D} pairs) as the controlling step in the melting process at these conditions and also observed \defect{5+7} defects as part of the melting mechanism.

In this paper we follow a similar approach to the one presented in \preonlineref\onlinecite{Mochizuki2013} where an ensemble of trajectories is generated and the effect of defects is investigated. However, due to the lower degree of superheating the critical step in the nucleation process is the formation of a liquid nucleus of sufficient size. At the temperatures we consider, the rate of formation of such critical nuclei is much lower than the rate of melting observed in \preonlineref\onlinecite{Mochizuki2013}. As a result, the required number of melting trajectories cannot be feasibly obtained from MD simulations simply by waiting for spontaneous melting events to occur. In this paper we generate unbiased melting trajectories based on the expectation of a nucleation-growth mechanism near phase coexistence, yielding hundreds of statistically independent samples. With the help of these trajectories we then assemble a detailed, time-resolved picture of the dynamic pathways that lead from a frozen crystal to a liquid nucleus at different degrees of superheating. This analysis yields a comprehensive picture of the roles different defects play in the mechanism of homogeneous melting as a function of temperature.

The remainder of the paper is organized as follows: in \presec\ref{sec:known_knowns} we introduce the types of defects that we refer to throughout the paper. In \presec\ref{sec:methods} we lay out our simulation methodology and in \presec\ref{sec:results} we present the results of our simulations. \Presec\ref{sec:discussion} summarizes and discusses our findings.

\section{Ice defects}\label{sec:known_knowns}
As part of our analysis we classify different point defect types that can occur in ice. Here we briefly introduce the types of defects we consider and summarize previous results on the role they play in melting. Higher dimensional defects such as dislocations and disclinations do not form spontaneously in the simulations presented in this paper and, hence, we limit this discussion to point defects only.

In order to discuss defects within the ice Ih structure it is instructive to first reiterate some of the properties of hexagonal ice\precite\cite{DeKoning2008}: (1) the water molecules (H$_2$O) are laid out in a wurtzite structure\precite\cite{Pauling1935,petrenko1999physics}; (2a) each molecule takes part in four hydrogen bonds to its nearest neighbors; (2b) two of the hydrogen bonds are donated to other molecules. Observations (2a) and (2b) together are known as the Bernal-Fowler ice rules\cite{Bernal1933,Pauling1935}. (3) The hydrogen bond network of a defect-free Ice Ih crystal can be decomposed into an array of 6-membered rings. However, the direction of hydrogen bonds in the bond network is not uniquely determined and, consequently, there is no long range proton order in Ice Ih\precite\cite{Pauling1935,petrenko1999physics}.
\begin{figure}[tbp]
  \begin{center}
    \includegraphics[width=1.0\columnwidth]{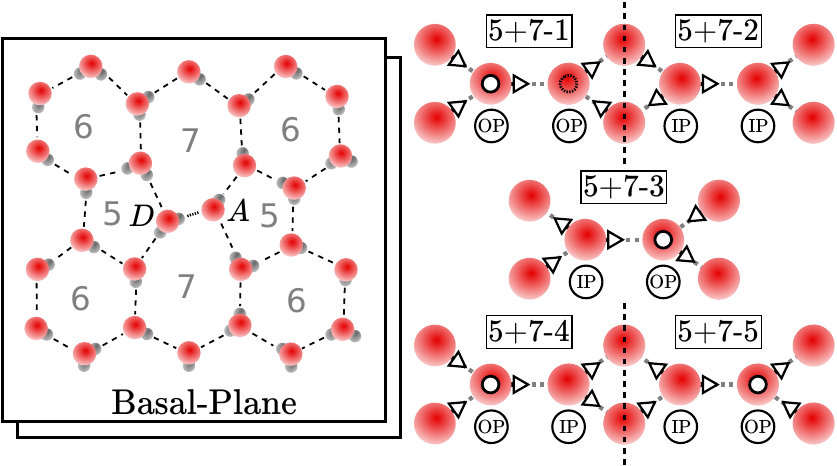}
  \end{center}
  \caption{Sketches of 5+7 defects. Red circles indicate oxygen atoms. \figleft 5+7 defect in the basal plane of a hexagonal ice crystal. $D$ and $A$ mark the central molecules of the 5+7 defect that form a hydrogen bond that is part of two 7-rings. Numbers indicate how many molecules comprise the rings that enclose them. \figright Classification of different 5+7 defect types according to the local arrangement of hydrogen bonds as defined in \preonlineref\onlinecite{Grishina2004} and used in this paper. White arrows point from donor to acceptor oxygen in hydrogen bonds. The circles on oxygens indicate hydrogen bonds that are formed along an orthogonal plane. IP and OP stands for in-plane and out-of-plane molecules, respectively, and indicates whether the plane that is spanned by the H-O-H triangle is parallel to the basal plane or orthogonal to it.}\label{fig:ice_hbonds_sketch_main}
\end{figure}

Breaking rule (1), i.e. displacing a molecule far from its location in the perfect lattice, leads to an interstitial-vacancy (\defect{I-V}) pair. It has been shown that translational diffusion within ice occurs by the movement of whole molecules\precite\cite{petrenko1999physics,Geil2005} and that the concentration of ionic defects is low compared to molecular defects\precite\cite{Eigen1958,DeKoning2008}. Hence, we expect that whole molecules also form the majority of \defect{I-V} defects in the lattice (and not ionic defects). In our simulations we only consider \defect{I-V} defects where whole molecules are moved out of their lattice position. No ionic defects can occur in our simulations.

Breaking condition (2) while keeping condition (1) intact leads to so-called Bjerrum- or \defect{L-D} pairs\precite\cite{Bjerrum1952}. \defect{L} and \defect{D} defects are, respectively, characterized by a missing or an excess proton within a small region of space so that the ice rules are not satisfied and can not be satisfied until a matching defect of the other type is encountered. Nevertheless, \defect{L-D} pairs are often found bound to each other forming an \defect{L+D} complex that exhibits a lower potential energy than a separated \defect{L-D} pair\precite\cite{Grishina2004}. In this work, we do not distinguish between separated \defect{L-D} pairs and \defect{L+D} complexes and call all structures where the ice rules are broken \defect{L-D} defects.

\defect{I-V} and \defect{L-D} defects will be referred to as \emph{mobile defects} throughout this work because, once a pair of these defects is formed and separated from each other, their movement through the system is relatively facile\precite\cite{Podeszwa1999a}. \preonlineref\onlinecite{Mochizuki2013} discusses the critical role of mobile defect pairs in melting under high superheating conditions, where the separation of a mobile defect pair is found to drastically lower the free energy barrier that needs to be overcome in order for melting to occur.

Lastly, breaking of rule (3) while keeping rule (2) intact constitutes another class of defects, which (in line with \preonlineref\onlinecite{Donadio2005}) we will call \emph{topological defects}. These are defects where the 6-ring structure of the perfect lattice is broken and instead there are other ring combinations present. Note that the ice rules are still fulfilled in these defects and that the molecules that take part are only shifted by small distances from their positions in the perfect lattice. The most prominent of these defects are so-called \defect{5+7} defects, first found in simulations by \textcite{Tanaka2002} and described in detail by \textcite{Grishina2004}. In \defect{5+7} defects two 5- and two 7-membered rings are found neighboring each other. These defects can then be further classified into different types according to the placement of protons around the central bond of the \defect{5+7} defect (see \prefig\ref{fig:ice_hbonds_sketch_main}) and according to the crystal plane they are formed in. \defect{5+7} defects are readily observed in equilibrium simulations that use the TIP4P family of water models\precite\cite{Tanaka2002,Kolafa2010a}.

\begin{figure}[tbp]
  \begin{center}
    \includegraphics[width=1.0\columnwidth]{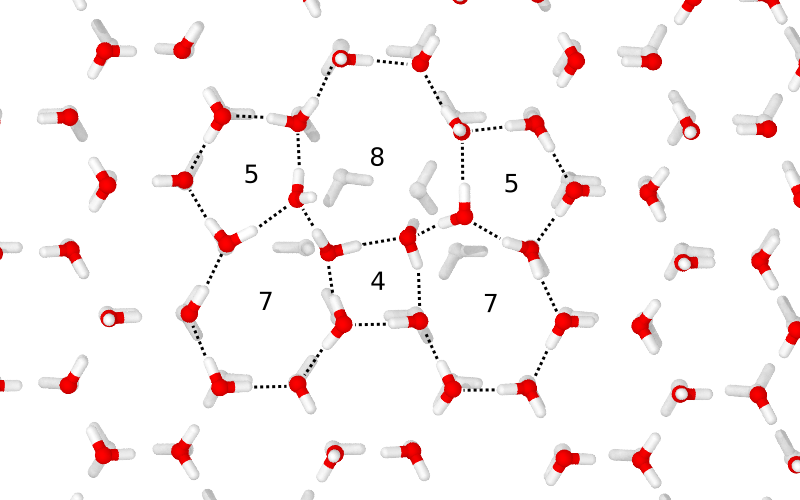}
  \end{center}
  \caption{Snapshot of a 455778 defect that is embedded into the basal plane of an Ice Ih crystal. The numbers mark the 4-, 5-, 7- and 8-membered rings in the H-bond network that make up the defect. The white molecules are located in an adjacent, defect free plane.}\label{fig:defect_455778}
\end{figure}

Other combinations of ring sizes are also possible: for example we encounter \defect{455778} defects that center around a 4-ring while still satisfying the ice rules (see \prefig\ref{fig:defect_455778}). Larger defect structures (such as the ones observed in \preonlineref\onlinecite{Donadio2005}) are frequently observed in melting trajectories and we will subsume all of these structures under the name \emph{extended topological defect} or \defect{E} defect. All topological defects have in common that there is no efficient mechanism for these defects to move which is why we will refer to them as \emph{immobile defects} in the following.

In the next section we present the simulation methodology used to generate trajectories that are then analyzed in terms of the defects that occur at various stages of melting.

\section{Methods}\label{sec:methods}
\subsection{Generating parts of reactive trajectories}\label{sec:gen_melt_traj}
\begin{figure}[tbp]
  \begin{center}
    \includegraphics[width=1.0\columnwidth]{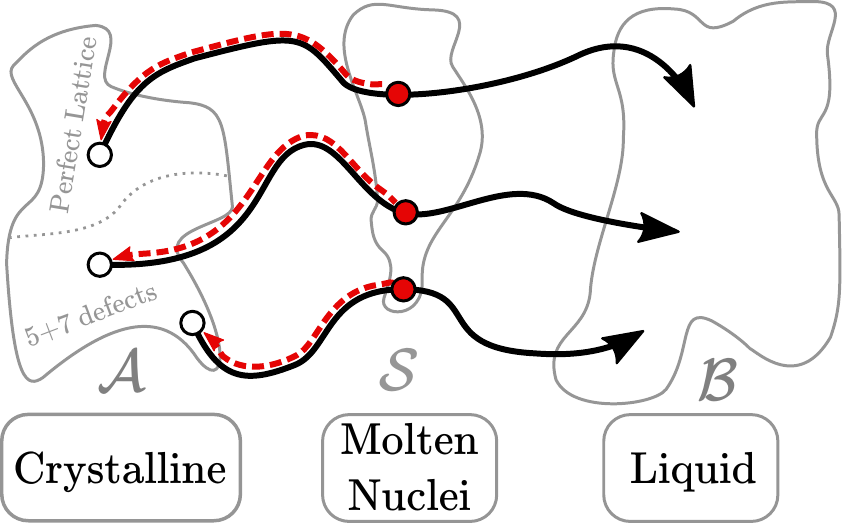}
  \end{center}
  \caption{Schematic representation of the method used to generate trajectories. Region $\mathcal{A}$ contains configurations that are completely solid with possibly a few \defect{5+7} defects remaining. Region $\mathcal{B}$ contains configurations that are liquid and region $\sts$ contains molten nuclei prepared by the procedure laid out in \presec\ref{sec:gen_melt_traj}. Our aim is to generate a sample of the early stages of melting trajectories (black curves with arrows). To do so, we integrate trajectories starting from configurations in $\sts$ until they reach $\sta$ (red, dashed arrows) and subsequently invert the direction of time. Trajectories that reach $\stb$ before $\sta$ are discarded.}\label{fig:trajectories_sketch}
\end{figure}

Our aim is to investigate the melting transition starting from configurations that contain an Ice Ih crystal with possibly a few \defect{5+7} defects (state $\sta$ in \prefig\ref{fig:trajectories_sketch}) and ending in the liquid state (state $\stb$). In particular, we are interested in the initial stages of these trajectories up to a state where a liquid nucleus has formed (state $\sts$ in \prefig\ref{fig:trajectories_sketch}). Note that we include configurations that contain \defect{5+7} defects into the definition of state $\mathcal{A}$. Such defect states occur readily in equilibrium trajectories under the conditions considered in this paper where the Ice Ih crystal is metastable with respect to the liquid state.

Under these conditions the two states $\sta$ and $\stb$ are separated by a free energy barrier so that the melting transition is a rare event. This means that the average waiting time $\tau_{\sta\stb}$ between preparing a system in an equilibrium frozen state in $\sta$ and a melting event that leads the system from state $\sta$ to the liquid state $\stb$ exceeds the timescale of relaxation in state $\mathcal{A}$, $\tau_\mathcal{A}$, by orders of magnitude:
\begin{equation}
  \begin{aligned}
  \tau_\mathcal{A} & \ll \tau_{\sta\stb}
  \end{aligned}
\end{equation}

This so-called \emph{separation of timescales} guarantees that the way melting events occur does not depend on the details of how frozen configurations are prepared and that, instead, we can think of the two states as being connected by an ensemble of melting trajectories that leave $\sta$ and end in $\stb$ without visiting $\sta$ in the meantime. This ensemble of trajectories is called the \emph{transition path ensemble}\precite\cite{Dellago1998,Bolhuis1998} (indicated by black arrows in \prefig\ref{fig:trajectories_sketch}).
\begin{figure}[tbp]
  \begin{center}
    \includegraphics[width=0.9\columnwidth]{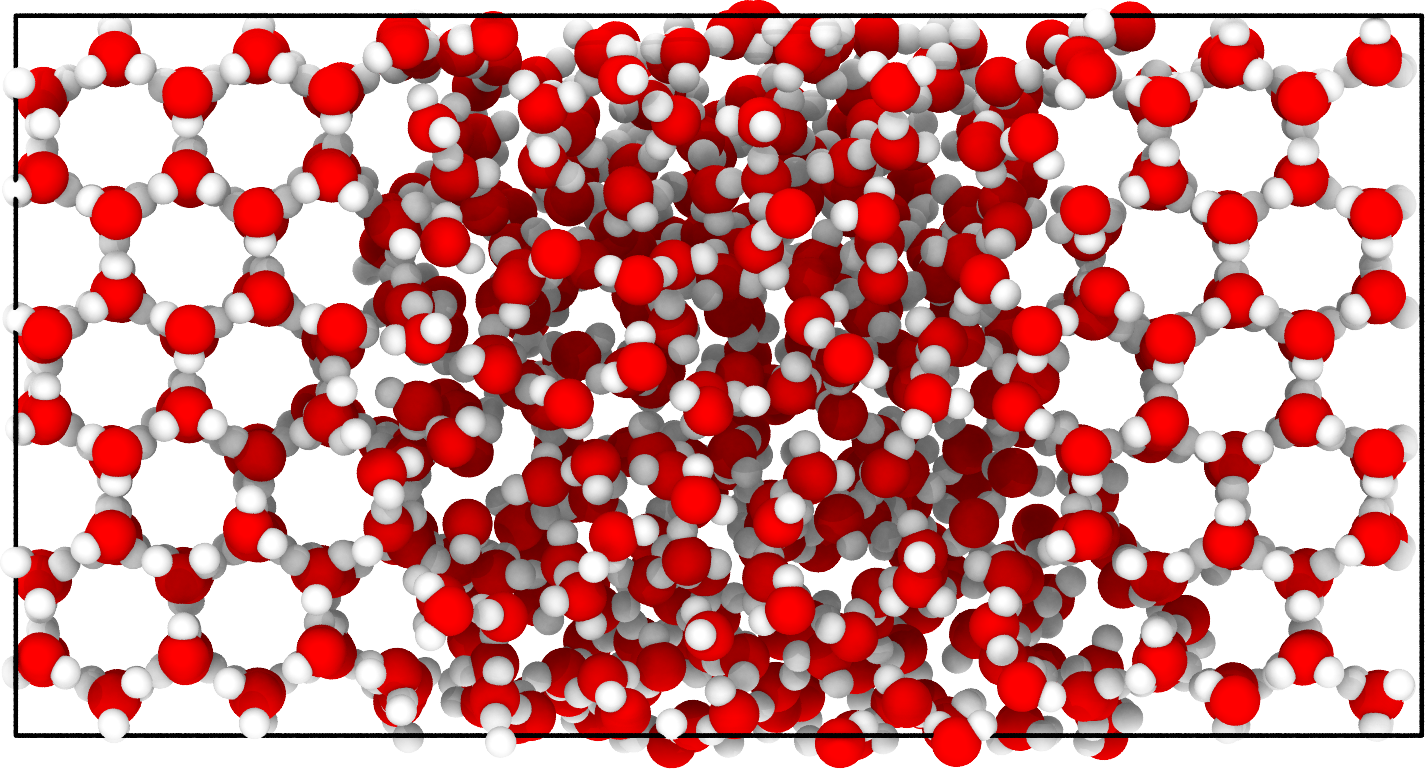}
  \end{center}
  \caption{Example of a configuration that contains a slab of liquid that consists of half the molecules in a 720-molecule water system. For clarity the potential energy of the configuration has been locally minimized.}\label{fig:example_slab} 
\end{figure}

To generate the initial parts of melting trajectories we take the following approach: \begin{enumerate}
  \item Pick a sample of configurations from equilibrium simulations of ice Ih performed at the chosen temperature and pressure.
  \item Construct a liquid domain inside this configuration and locally equilibrate the resulting configuration. The resulting ensemble is denoted with $\sts$.
  \item Run molecular dynamics simulations using a symplectic integration scheme starting from $\sts$ until state $\sta$ or $\stb$ is reached. The resulting trajectories are referred to as \emph{backwards trajectories}.
  \item Invert the time direction of the backwards trajectories that end in $\sta$ to get a sample of trajectories that lead from $\sta$ to $\sts$.
\end{enumerate}

The use of a symplectic integration scheme guarantees that the sample of trajectories obtained by integrating backwards in time has the same statistics as the ensemble of trajectories that leads from $\sta$ to $\sts$ when integrating forward\precite\cite{Dellago2002a}.

Similar to the so-called seeding method\precite\cite{Bai2005,Bai2006,Espinosa2014a,Espinosa2016b} we choose the liquid clusters to be spherical in shape. In reality, the shape of the liquid nuclei that form during homogeneous melting is likely not perfectly spherical due to slight differences in the interface tension associated with different crystal planes\precite\cite{Handel2008,Davidchack2012,Espinosa2014b,Espinosa2016c} as well as due to the different dynamics of crystal growth along the plane normals\precite\cite{Nada1997,Nada2005a,Rozmanov2012,Espinosa2016d}. To assess the effect different cluster shapes have on the early stages of melting trajectories, we perform additional simulations at the ice-liquid coexistence temperature that start from configurations with a slab shaped liquid domain such as the one shown in \prefig\ref{fig:example_slab}.

We refer to trajectories that are constructed from spherically shaped liquid domains as \emph{spherical-geometry} trajectories and to the ones constructed from slab shaped liquid domains as \emph{slab-geometry} trajectories.

\subsection{Simulation details}
The simulations presented in this paper are obtained using the TIP4P/Ice water model\precite\cite{Abascal2005a} with a time reversible and symplectic rigid body integration scheme\cite{Miller2002,Kamberaj2005} as implemented in the LAMMPS simulation package\cite{Plimpton1995}\footnote{LAMMPS version (Aug 22 2018) has been used to generate trajectories.}. Thermo- and barostats are implemented using Nos\'{e}-Hoover chains\cite{Hoover1985a,martyna1996explicit,Martyna1992} where the $x$, $y$ and $z$ directions are independently barostatted to a pressure of \SI{1}{\bar}. Long-range interactions are treated using a particle-mesh Ewald method (PPPM\cite{Hockney1988,Darden1993}) with an accuracy of \num{e-4} and the timestep is set to \SI{1}{\femto\second}. Snapshots are saved for analysis every \SI{10}{\pico\second}.

\subsubsection*{Spherical-geometry trajectories}
MD simulations used to generate spherical geometry trajectories are carried out with 2880 molecules in an almost cubic simulation box (see \prefig\ref{fig:example_liquid_cluster}). The initial dimensions of the boxes are $44.9 \times 46.7 \times 44.0 \,\SI{}{\angstrom\cubed}$ in the directions orthogonal to the secondary-prism plane, the prism plane and the basal plane, respectively. To generate initial configurations for the backward trajectories we follow the following procedure:
\begin{enumerate}
  \item  Construct a proton ordered Ice XI lattice and randomly reorder the hydrogen bonds using a Monte Carlo procedure\precite\cite{Ayala2003} in order to obtain ice Ih. Here we require that the total dipole moment of the configuration is zero at the end of the procedure.
  \item Equilibrate these configurations in a $\SI{30}{\nano\second}$ parallel tempering\precite\cite{Lyubartsev1992,Marinari1992,Okabe2001,Mori2010} trajectory using replicas starting from a temperature of $\SI{258}{\kelvin}$ up to and including $\SI{328}{\kelvin}$ spaced \SI{5}{\kelvin} apart.
  \item Pick a sample of configurations from the parallel tempering simulation at the desired temperature.
  \item Pick a random center for the liquid nucleus in each of the configurations and find the molecules within a radius of \SI{15}{\angstrom}.
  \item Heat the selected molecules using a thermostat while keeping the molecules outside the sphere fixed until the crystal structure in the selected region breaks down.
  \item Equilibrate at the target temperature by first keeping the molten fraction fixed and propagating the molecules in the crystalline phase (for \SI{20}{\pico\second}) and then keeping the crystalline molecules fixed and propagating the molten molecules (for \SI{25}{\pico\second}). This procedure hinders the molecules from recrystallizing because the molecules inside and outside of the selected volume can not collectively reorder into a frozen configuration.
\end{enumerate}

\subsubsection*{Slab-geometry trajectories}
Slab-geometry trajectories are generated using 720 molecules in an elongated box with initial dimensions $44.9 \times 23.3 \times 22.0 \,\SI{}{\angstrom\cubed}$ (see \prefig\ref{fig:example_slab} for an example configuration). The volume that contains the liquid is chosen so that the solid-liquid interfaces are parallel to the secondary prism face of the Ice Ih crystal. This is the geometry that has been found to preferentially form in \preonlineref\onlinecite{Keyes2015}. No initial equilibration is performed because the system has sufficient time to equilibrate before the slab collapses into one of the two competing phases.

The resulting trajectories are quite different from the spherical-geometry trajectories at the same temperature of \SI{268}{\kelvin}. The spherical liquid domains shrink rapidly and predictably due to surface tension. For the slab geometry, periodic boundary conditions eliminate contributions from the surfaces where the liquid domain wraps around the simulation box\precite\cite{Leung1990,Troster2005}. Surface tension therefore does not drive the growth of the slab-shaped liquid domains because a change in the volume of the liquid domain has no effect on the overall liquid-solid surface area.

The temperature of $T = \SI{268}{\kelvin}$ was chosen so that roughly half of the trajectories started with a slab-shaped domain melt and the others freeze, further reducing the thermodynamic force that drives the growth and shrinkage of the liquid domain. This temperature is slightly below the melting temperature at \SI{1}{\bar} of \SI{272.2}{\kelvin} reported by \textcite{Abascal2005a}; a discrepancy that is likely due to the non-negligible influence of the solid-liquid interfaces in the comparatively small simulation box with 720 molecules. The result are trajectories where the two solid-liquid interfaces that delimit the liquid domain diffuse in the direction of their surface normal until they are close enough that a fluctuation in the shape of the interfaces brings them into contact. The distance between the interfaces at which this contact occurs is small, roughly one layer of 6-rings or \SI{7}{\angstrom}. 

A larger simulation cell would yield a larger, and more realistic, contact distance. Increasing this distance considerably, however, comes at great computational cost, for two reasons: first, the contact distance scales logarithmically with the size of the simulation box\precite\cite{Moritz2020}. Secondly, increasing the size of the simulation box slows down the diffusion of the two interfaces which in turn sharply increases the length of the required trajectories. The system size with interface areas of $23.3 \times 22.0 \,\SI{}{\angstrom\squared}$ was chosen as a compromise between maximizing the contact distance between the two interfaces and minimizing the simulation time required to obtain trajectories.

\subsubsection*{Gathered data}
In total, three sets of trajectories leading from the prepared configurations $\sts$ to the frozen state $\sta$ were generated: $108$ slab-geometry trajectories at a temperature of $T = \SI{268}{\kelvin}$, $504$ spherical-geometry trajectories at $T = \SI{268}{\kelvin}$, and $448$ spherical-geometry trajectories at $T = \SI{303}{\kelvin}$. 

The temperature of $\SI{303}{\kelvin}$ is $11\%$ superheated relative to the melting point\precite\cite{Abascal2005a} and has been chosen so that the spherical nuclei with a radius of $\SI{15}{\angstrom}$ are slightly subcritical. Out of the $504$ trajectories that were run we observed 56 trajectories that melted before they could reach the frozen state $\sta$. On average $\mathcal{A}$ is reached in \SI{13.6}{\nano\second} (slab, \SI{268}{\kelvin}), \SI{5.1}{\nano\second} (spherical, \SI{268}{\kelvin}), and \SI{2.3}{\nano\second} (spherical, \SI{303}{\kelvin}).

Data on equilibrium properties presented in this paper were obtained from the parallel tempering simulations that were also used to generate initial configuration for the backward simulation runs. The data from different replicas is combined using the weighted histogram analysis method (WHAM\precite\cite{Ferrenberg1989,Kumar1992}) as implemented in the PyEMMA package\precite\cite{Scherer2015}. 

We now invert the direction of time in these generated trajectories, which become examples of the early stages of melting. We also set $t=0$ at the crystalline endpoint of each trajectory, so that the system adopts a frozen configuration at time zero and melting commences as time increases. These time conventions will be used throughout the remainder of the paper.

\subsection{Defect detection}
To detect defects in the hexagonal ice structure we adapt an array of different techniques previously used to investigate the properties of water and ice. These include, after locally minimizing the potential energy of a configuration, an analysis of the hydrogen bond network\precite\cite{Mochizuki2013,VilaVerde2012}, in particular the ring structures found therein\precite\cite{Donadio2005}, as well as the analysis of configurations relative to a reference configuration to facilitate the detection of molecular interstitials and vacancies\precite\cite{Mochizuki2013}. See \preapp\ref{app:defect_detection_details} for a detailed description of the algorithms used.
\begin{figure}[tbp]
  \begin{center}
    \includegraphics[width=1.0\columnwidth]{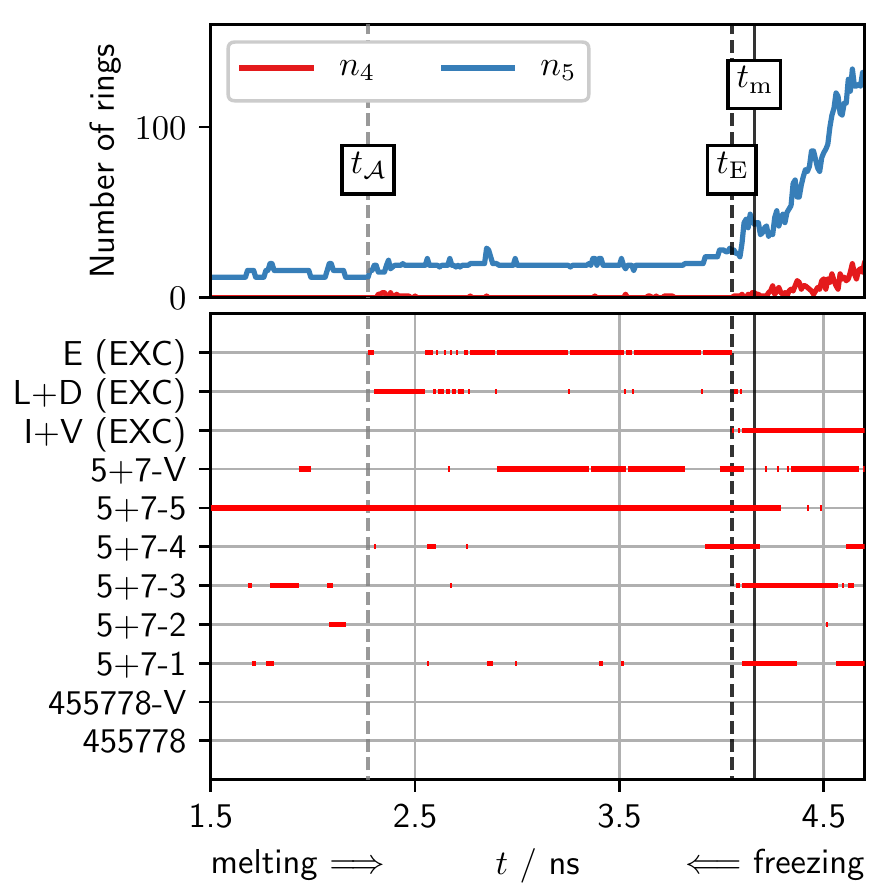}
  \end{center}
  \caption{Analysis of an example melting trajectory. At the beginning of the chosen timeframe a number of 5+7-5 defects are present in an otherwise frozen configuration. The vertical lines indicate the times used to split the trajectory into the three stages decribed in \presec\ref{sec:results}. Red markings in the lower half of the plot indicate the presence of at least one defect of the given type in the system. Due to the algorithm used only one of the defect types marked with \defect{(EXC)} can be detected at a time (see \preapp\ref{app:defect_detection_details}).}\label{fig:example_traj_analysis}
\end{figure}

\begin{figure}[tbp]
  \begin{center}
  \includegraphics[width=0.90\columnwidth]{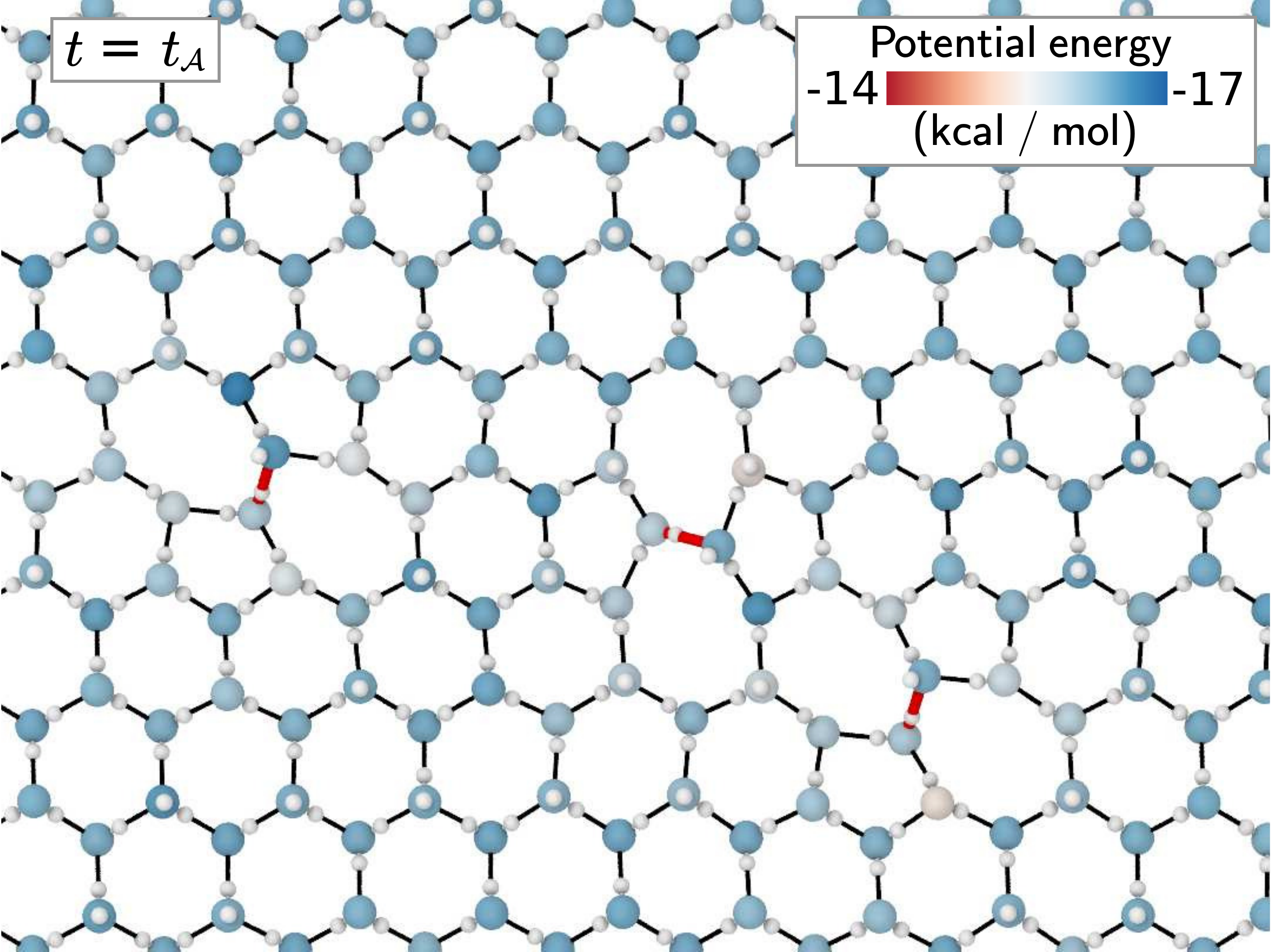} \\
  \vspace{.5cm}
  \includegraphics[width=0.90\columnwidth]{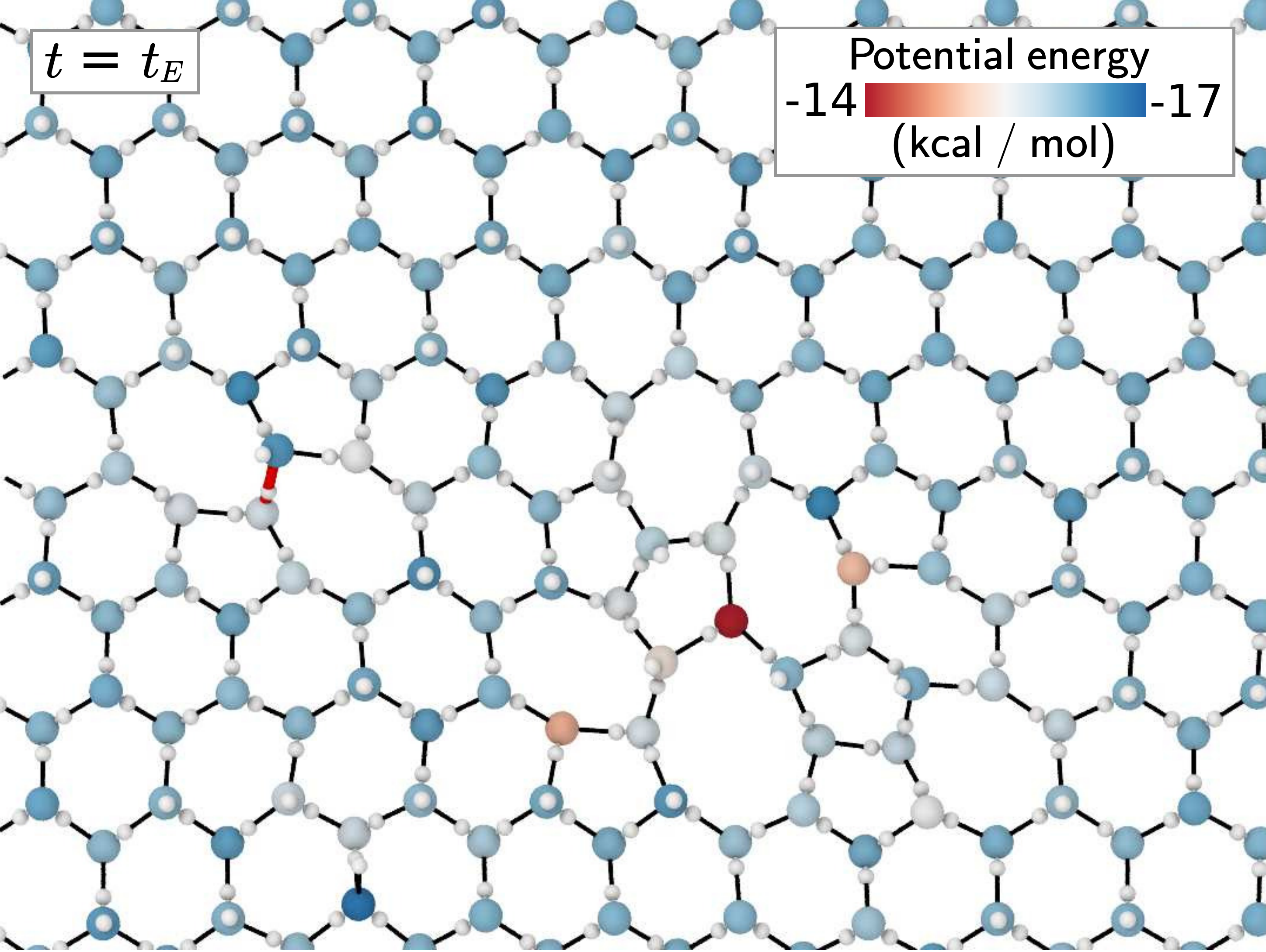}
  \end{center}
  \caption{Snapshots of the trajectory analyzed in \prefig\ref{fig:example_traj_analysis} at time $\ta$ (top) and at time $\te$ (bottom). Shown is a single basal plane that contains the 5+7 defects present at time $\ta$ (red, thick bonds). The colors of molecules indicate their contribution to the total potential energy. The black lines indicate hydrogen bonds between molecules.}\label{fig:example_traj_snapshots}
\end{figure}

These techniques yield an analysis like the one shown in \prefigs\ref{fig:example_traj_analysis} and \ref{fig:example_traj_snapshots}. It includes counts of the number of 4- and 5- membered rings ($n_4(t)$ and $n_5(t)$, respectively) that are roughly proportional to the size of the liquid domain. It also indicates whether certain defects have been detected in the configuration. The defect types distinguished are \defect{I-V} pairs, \defect{L-D} pairs, and topological defects in the H-bond network: \defect{5+7} defects, \defect{455778} defects, and extended topological defect structures where the ice rules are fulfilled (\defect{E} defects). The \defect{5+7} defects are further split into the five horizontal types\precite\cite{Grishina2004} (\defect{5+7-1} through \defect{5+7-5}) that are formed within the basal plane of the lattice and the vertical type (\defect{5+7-V}) where the \defect{5+7} defect is formed in a plane orthogonal to the basal plane.

Note that the algorithm used to detect defects of type \defect{I-V}, \defect{L-D}, and \defect{E} does so in an exclusive fashion: if an \defect{I-V} pair is present, \defect{L-D} and \defect{E} defects can not be detected and if an \defect{L-D} pair is present, \defect{E} defects can not be detected (see \preapp\ref{app:defect_detection_details}, \prefig\ref{fig:defect_detection_flow_chart}).

As part of the analysis we determine three times along each melting trajectory: $\ta$, $t_\mathrm{E}$, and $t_{\mathrm{m}}$:
\begin{itemize}
  \item $\ta$ is the time when the system leaves state $\mathcal{A}$ for the last time, i.e. it is the last time where there are only 5+7 or 455778 defects present in the system.
  \item $\te$ is the last time when no mobile defects are present in the system. In the time between $\ta$ and $\te$, \defect{E} defects are present in the system; L-D and I-V pairs may also form during this interval, but by definition they must recombine before time $\te$. If no \defect{E} defect occurs along a trajectory, then $\ta$ equals $\te$.
  \item $\tm$ is the time when the extended liquid domain forms. This event is associated with an increase in the rate at which 5-membered rings appear, i.e., the slope of $n_5(t)$. Because this increase occurs on top of significant background fluctuations in $n_5(t)$, we look for an increase of $n_5(t)$ over a timespan of \SI{200}{\pico\second} that is larger than a given threshold $\Delta n_5$ (see \pretab\ref{tab:n5_thresholds}).
\end{itemize}

\begin{table}[tb]
  \centering
  \caption{Thresholds in the growth of the number of 5-rings over a \SI{200}{\pico\second} period, $\Delta n_5$, used to detect time $\tm$.}\label{tab:n5_thresholds}
  \begin{ruledtabular}
    \begin{tabular}{lcc} 
      Geometry & Temperature ($T$) & $ \Delta n_5$ \\
      \hline
      Slab   & \SI{268}{\kelvin} & 8 \\
      Sphere & \SI{268}{\kelvin} & 8 \\
      Sphere & \SI{303}{\kelvin} & 15 \\
    \end{tabular}
  \end{ruledtabular}

\end{table}

\section{The three stages of melting}\label{sec:results}
\begin{figure}[tbp]
  \begin{center}
    \includegraphics[width=1.0\columnwidth]{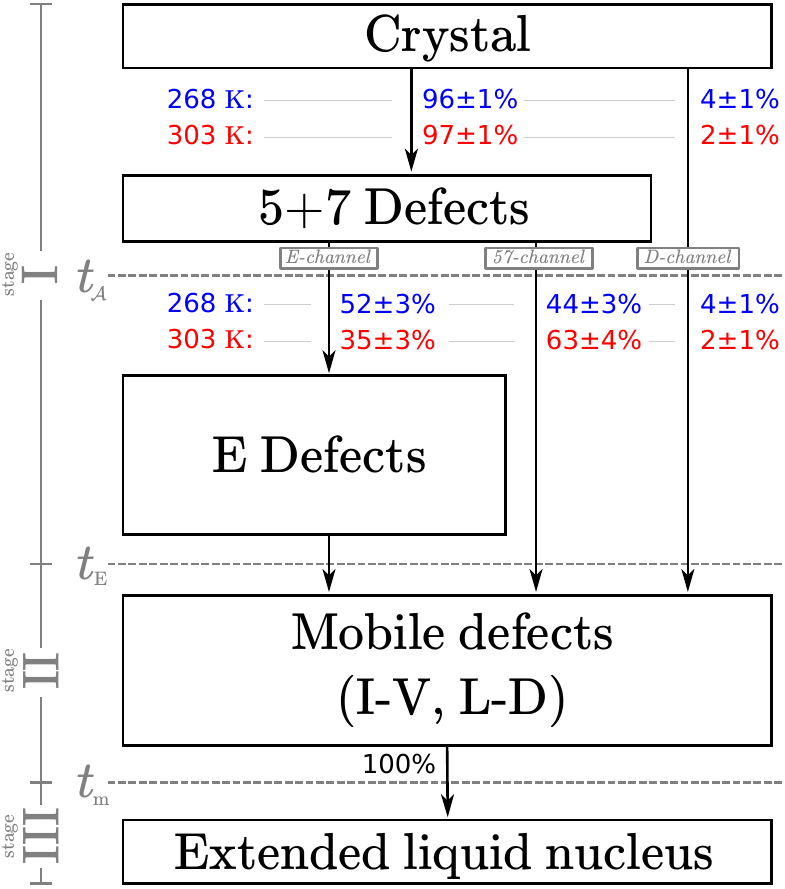}
  \end{center}
  \caption{Diagram of the stages observed during the melting of hexagonal ice crystals and the pathways that are taken by trajectories in which a spherical liquid nucleus forms. The percentages represent the fractions of trajectories that proceed along the indicated pathway relative to the total number of melting trajectories.}\label{fig:reaction_channels_sketch}
\end{figure}

Based on the times defined in the previous section we now define three stages of the melting mechanism (cf.\,\prefig\ref{fig:reaction_channels_sketch}): stage I ($0 \leq t < \te$) where only immobile defects are present in the system, stage II ($\te \leq t < \tm$) where one or more mobile defects have formed and stage III ($t \geq \tm$) where an extended liquid nucleus has formed. In the following sections we first discuss the spherical-geometry trajectories. \Presec\ref{sec:differences_slab} then discusses the differences found in slab-geometry trajectories.

\subsection{Stage I: topological defects ($t \le \te$)}
\begin{figure}[tbp]
  \begin{center}
    \includegraphics[width=0.95\columnwidth]{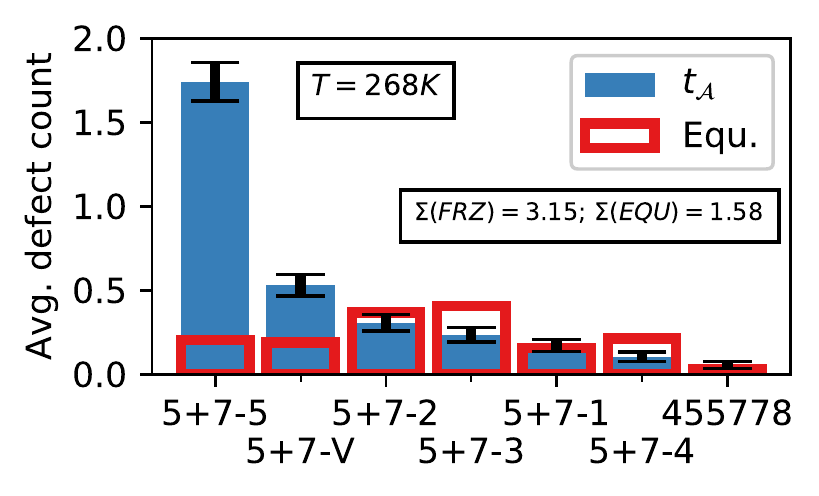}
    \includegraphics[width=0.95\columnwidth]{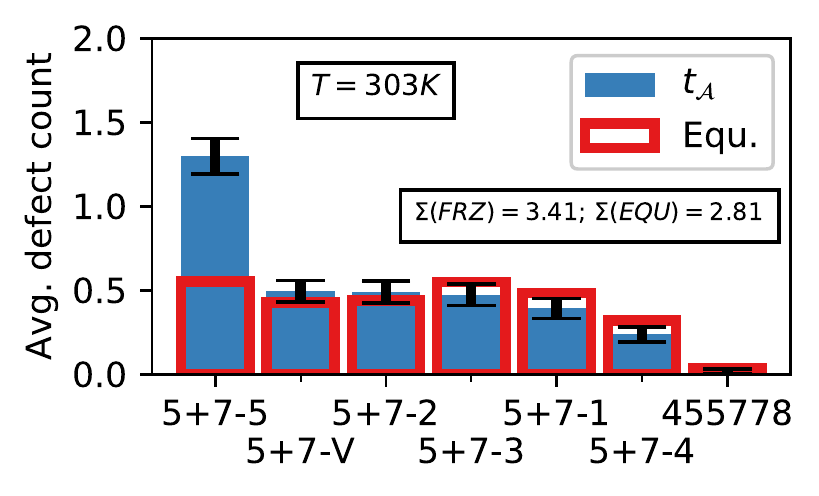}
  \end{center}
  \caption{Average number of 5+7 defects found in a cubic simulation box with 2880 molecules in equilibrium, alongside the probability of finding them in melting trajectories at time $\ta$. Results are shown in the top panel for $T = \SI{268}{\kelvin}$ and in the bottom panel for $T = \SI{268}{\kelvin}$. The black error bars indicate a confidence level of $95\%$. $\Sigma(\text{FRZ})$ and $\Sigma(\text{EQU}) $ indicate the average overall defect counts at time $\ta$ and in equilibrium, respectively.}\label{fig:defect_probs_comp_equ}
\end{figure}

In equilibrium as well as in stage I of melting trajectories, the bulk of defects that are present are of type \defect{5+7}. A prominent role in melting is played by \defect{5+7-5} defects as can be seen in \prefig\ref{fig:defect_probs_comp_equ} where we report the average numbers of defects found at time $\ta$ compared to the average numbers found in equilibrium. The difference in defect numbers between these two scenarios is listed in \pretab\ref{tab:equ_ta_defect_abundances}.
\begin{table}[tb]
    \centering
    \caption{Average numbers of defects $\left<n\right>$ at time $\ta$ and in equilibrium, as well as the difference $\Delta$ between the two numbers, for different defect types and at different temperatures $T$. Reported errors are $95\%$ confidence intervals calculated assuming that the numbers of defects are Poisson distributed. This assumption is supported by an analysis of the defect statistics presented in \preapp\ref{app:additional}, \prefig\ref{fig:first_in_A_defect_stats_268K}.}\label{tab:equ_ta_defect_abundances}
    \begin{ruledtabular}
      \begin{tabular}{cccc} 
      T = \SI{268}{\kelvin}  & $\left<n\right>(t=\ta)$ & $\left<n\right> (\text{Equ.})$ & $\Delta$ \\
      5+7-5 & $1.74 \pm 0.12$ & $0.21$ & $ 1.53$ \\
      5+7-V & $0.53 \pm 0.06$ & $0.19$ & $ 0.34$ \\
      5+7-2 & $0.31 \pm 0.05$ & $0.37$ & $-0.06$ \\
      5+7-3 & $0.24 \pm 0.04$ & $0.41$ & $-0.17$ \\
      5+7-1 & $0.17 \pm 0.04$ & $0.16$ & $ 0.01$ \\
      5+7-4 & $0.11 \pm 0.03$ & $0.21$ & $-0.11$ \\
     455778 & $0.06 \pm 0.02$ & $0.03$ & $ 0.03$ \\
        SUM & $3.15 \pm 0.16$ & $1.58$ & $ 1.57$
    \end{tabular}
    \begin{tabular}{cccc} 
      T = \SI{303}{\kelvin} & $\left<n\right>(t=\ta)$ & $\left<n\right> (\text{Equ.})$ & $\Delta$ \\
      5+7-5 & $1.30 \pm 0.11$ & $0.55$ & $ 0.74$ \\
      5+7-V & $0.50 \pm 0.07$ & $0.43$ & $ 0.07$ \\
      5+7-2 & $0.49 \pm 0.06$ & $0.44$ & $ 0.05$ \\
      5+7-3 & $0.48 \pm 0.06$ & $0.55$ & $-0.08$ \\
      5+7-1 & $0.39 \pm 0.06$ & $0.48$ & $-0.09$ \\
      5+7-4 & $0.24 \pm 0.05$ & $0.32$ & $-0.08$ \\
     455778 & $0.02 \pm 0.01$ & $0.04$ & $-0.02$ \\
        SUM & $3.41 \pm 0.17$ & $2.81$ & $ 0.60$
    \end{tabular}
    \end{ruledtabular}
\end{table}

At both temperatures the overall excess of defects found at time $\ta$ is mostly accounted for by the excess of \defect{5+7-5} defects. At $T = \SI{268}{\kelvin}$ the number of \defect{5+7-V} defects is also significantly enhanced while the number of \defect{5+7-3} and \defect{5+7-4} defects is slightly reduced. At $T = \SI{303}{\kelvin}$ the average number of defects of type \defect{5+7-V} and types \defect{5+7-1} through \defect{5+7-4} are roughly equal to the numbers observed in equilibrium.

At time $\ta$ one of two things happens: either an \defect{E} defect forms (\emph{E-channel}) or a mobile defect forms directly (\emph{57-channel}). \Prefig\ref{fig:reaction_channels_sketch} shows the fractions of trajectories that pass through each of these channels. As the temperature decreases from \SI{303}{\kelvin} to \SI{268}{\kelvin} the number of trajectories that involve an \defect{E} defect increases from $35\%$ to $52\%$.

In order to assess the role of \defect{5+7} defects in the melting mechanism, we analyze the locations of defects relative to the center of the volume where the liquid nucleus eventually forms (the \emph{nucleus volume}), as well as relative to the sites where mobile defects form at time $\te$. To do so, we define the pair-correlation function between defects of reference type $S$ and defects of target type $T$,
\begin{equation}\label{equ:g_def}
  g_{ST}(r) = \frac{1}{c_S} \frac{H_{ST}(r, \Delta r)}{\rho_T V(r,\Delta r)},
\end{equation}
where $H_{ST}(r, \Delta r)$ is a histogram of all pairwise distances between defects of type $S$ and type $T$ observed in a set of configurations, $\rho_T$ is the equilibrium number density of defects of type $T$, and $c_S$ is the total number of defects of type $S$ observed. $V(r,\Delta r)$ is the volume of a spherical shell with inner radius $r - \Delta r/2$ and outer radius $r + \Delta r/2$, where $\Delta r$ is the bin width of the histogram. Note, that $g_{ST}(r)$ is not symmetric in $S$ and $T$ because $c_S \neq \rho_S$. Nevertheless, we expect $\lim_{r\to\infty} g_{ST}(r) = 1$.

We also define $n_{ST}(r)$ as the average number of defects $T$ within a sphere of radius $r$ centered on a defect of type $S$, i.e.
\begin{equation}
  n_ {ST}(r) =  \rho_T \sum_{r_i < r} V(r_i, \Delta r) \, g_{ST}(r_i).
\end{equation}
$g_T(r)$ and $n_T(r)$ are the analogous quantities where the reference points are the center of the spherical nucleus volume.
\begin{figure}[tbp]
  \begin{center}
    \includegraphics[width=1.0\columnwidth]{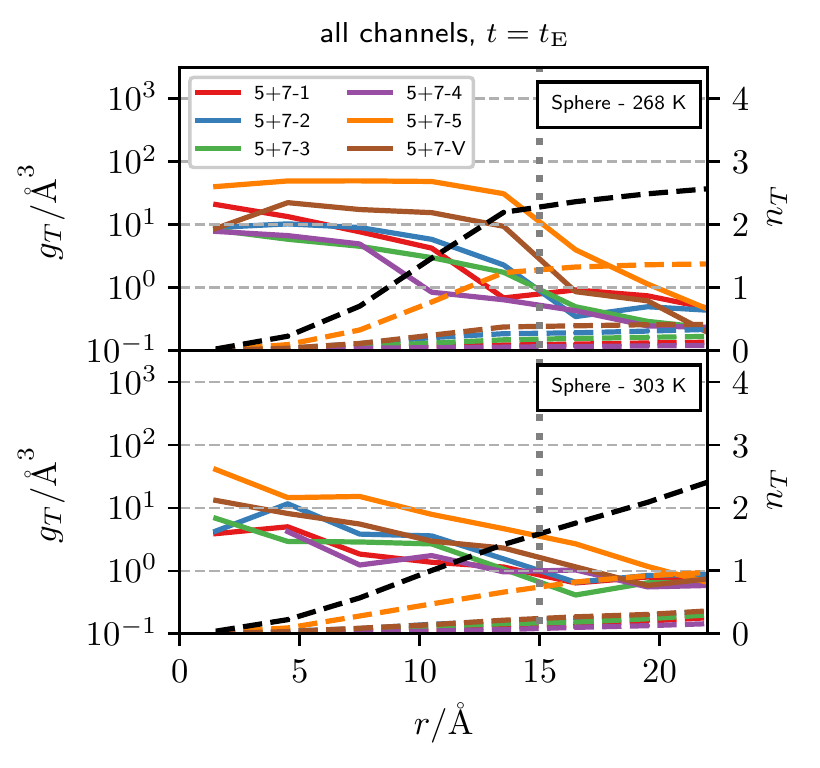}
  \end{center}
  \caption{Abundance of \defect{5+7} type defects as a function of distance $r$ from the center of the liquid nuclei that were used to seed the simulations. The gray vertical dashed line indicates the \SI{15}{\angstrom} radius of the liquid nucleus. Shown are the normalized defect densities $g_T(r)$ (solid) and the number $n_T(r)$ of defects within a spherical volume of radius $r$ around the center of the nascent liquid nucleus (dashed). The dashed black line indicates the sum of $n_T$ over all defect types, $T$.}\label{fig:melting_trajs_defect_center_i_v_T_268K_T_303K}
\end{figure}

\Prefig\ref{fig:melting_trajs_defect_center_i_v_T_268K_T_303K} shows $g_T(r)$ and $n_T(r)$ obtained from configurations observed at time $\te$ (including trajectories that proceed through both the 57- and E-channels). At $T = \SI{268}{\kelvin}$ there is up to a 50-fold excess over equilibrium in the density of \defect{5+7} defects inside the radius of the liquid bubble that forms later on. The most abundant defect type inside this volume is the \defect{5+7-5} defect with an average of $1.2$ defects found within the radius of \SI{15}{\angstrom} while the average total number of \defect{5+7} defects within this volume is $2.2$. The next most common defect within the nucleus volume are \defect{5+7-V} defects.

Under superheating conditions at $T = \SI{303}{\kelvin}$ the excess is slightly less pronounced. Nevertheless, on average we find $1.5$ \defect{5+7} defects within \SI{15}{\angstrom} of the center of the forming bubble (\defect{5+7-5} defects: $0.7$).
\begin{figure}[tbp]
  \begin{center}
    \includegraphics[width=1.0\columnwidth]{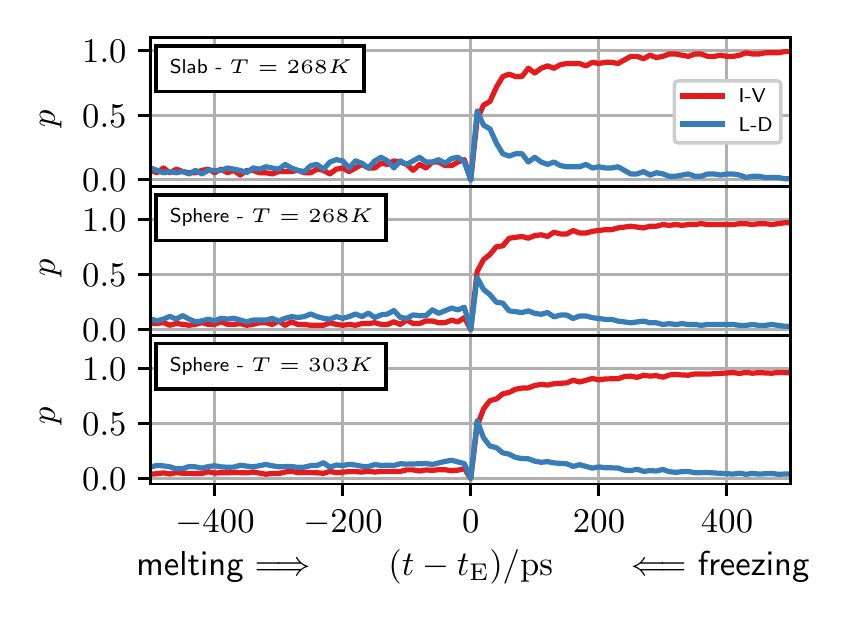}
  \end{center}
  \caption{Probabilities of finding the system with defects of type \defect{I-V} or \defect{L-D}, as functions of time. To calculate the time-dependent averages, trajectories have been aligned on time $\te$. The states shown are detected in a mutually exclusive fashion, i.e.\ if there exists at least one \defect{I+V} defect in the system, it is considered to be in the \defect{I+V} state regardless of the number of \defect{L+D} defects in the system. Notice the similarity between the data obtained with the slab and the spherical geometry; the probabilites are largely independent of the shape of cluster that is formed as well as of temperature. The probabilities of finding the system in state \defect{I+V} and \defect{L+D} by definition vanish at $\te$ and sum to one for $t > \te$.}\label{fig:defect_stats_aligned_on_last_s}
\end{figure}

At $T = \SI{268}{\kelvin}$ we observe a suppression of defects outside the nucleus volume relative to equilibrium. This suppression develops despite the fact that we used equilibrium configurations to seed the simulations, a procedure that enforces that the environment around the liquid domain is initially in equilibrium. The suppression of defect densities around the liquid nucleus indicates that the presence of the nucleus facilitates annealing of existing defects in the ice structure. We expect this effect to subside with increasing distance from the nucleus, however, the simulation box used in our simulations is not large enough to observe this return to average densities.

It is important to note here that precisely at coexistence the radius of the critical nucleus is macroscopically large, and that the nucleus radius of \SI{15}{\angstrom} used to seed our simulations was chosen for comparison with simulations performed under superheating. At the physically more realistic temperature of $\SI{303}{\kelvin}$ the nucleus size used to seed simulations is chosen close to the critical nucleus size at this temperature. This allows us to extrapolate these simulation results to larger system sizes. Notably, under these conditions there is no significant suppression of defect densities outside the nucleus volume.
\begin{figure}[tbp]
  \begin{center}
    \includegraphics[width=1.0\columnwidth]{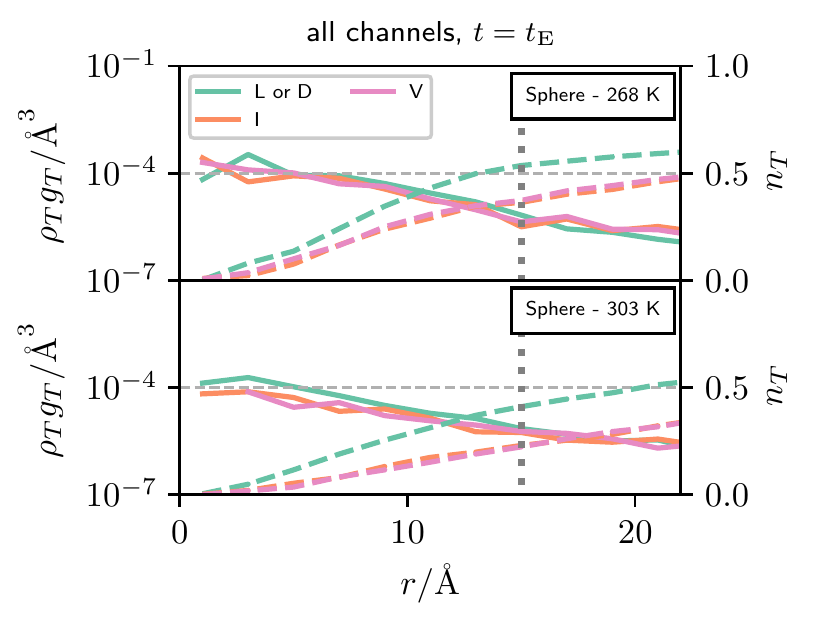}
  \end{center}
  \caption{Analysis of locations where \defect{L-D} pairs, and \defect{I} and \defect{V} defects form at time $\te$ relative to the center of the future liquid nucleus. Shown are the densities $\rho_T g_T(r)$ (solid) and the number of defects within a spherical volume of radius $r$ around the center, $n_T(r)$ (dashed). The gray dotted line indicates the \SI{15}{\angstrom} radius of the nucleus volume.} \label{fig:melting_trajs_defect_center_i_v}
\end{figure}

\subsection{Stage II: mobile defects ($\te < t \le \tm$)}
In this next stage, a mobile defect, i.e., either an \defect{L-D} or an \defect{I-V} pair has formed. Which defect type has formed as a function of time relative to $\te$ is shown in \prefig\ref{fig:defect_stats_aligned_on_last_s} for different temperatures and cluster geometries. Just over half of the configurations ($56\%$)  observed at temperature $T = \SI{303}{\kelvin}$ contain an \defect{L-D} pair, while the other ones contain an \defect{I-V} pair. The fraction of configurations where an \defect{I-V} pair is present then rapidly increases, reaching $98\%$ \SI{0.5}{\nano\second} later. Similar behavior can be observed at $T = \SI{268}{\kelvin}$ regardless of the geometry of the liquid nucleus, giving us confidence that the observed timescale of roughly \SI{0.5}{\nano\second} for the time between forming an \defect{L-D} pair and forming an \defect{I-V} pair is largely independent of the geometry of the nucleus that forms and of temperature.
\begin{figure}[tbp]
  \begin{center}
    \includegraphics[width=1.0\columnwidth]{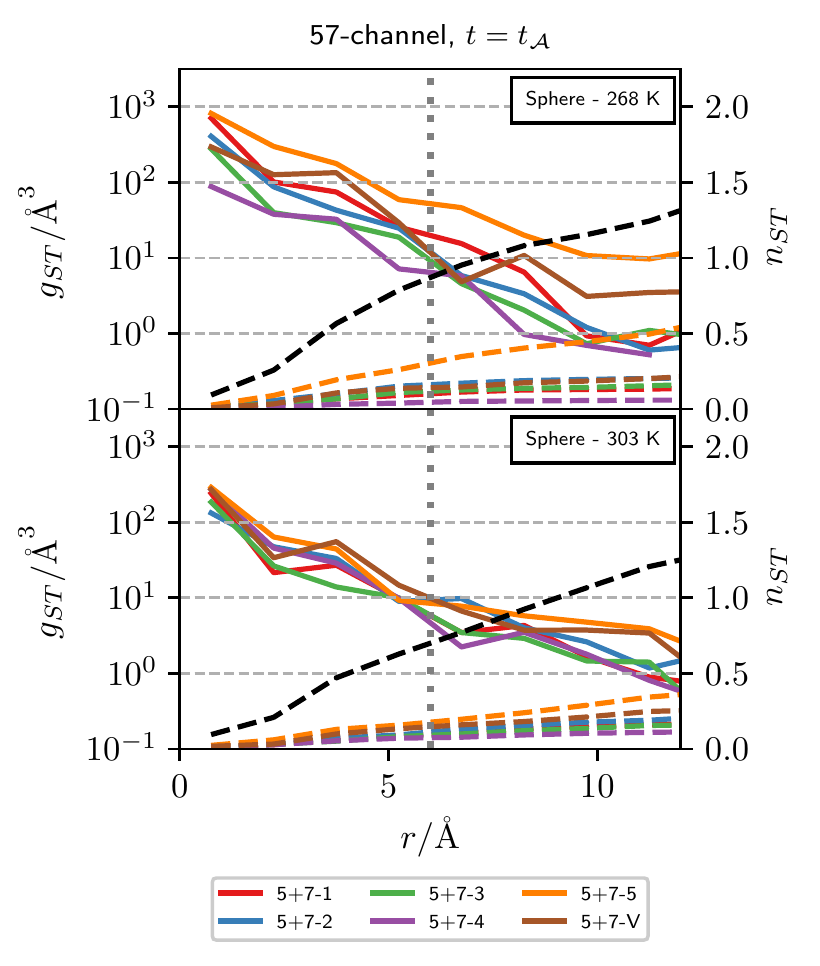}
  \end{center}
  \caption{Analysis of the positions of immobile defects around the site where a mobile defect pair of type \defect{I-V} or \defect{L-D} forms at time $\ta$. Shown are the pair-correlation functions $g_{ST}$ (solid) and the average numbers of defects within a sphere of radius $r$, $n_{ST}$ (dashed). Here, the reference defect type $S$ are the defect types \defect{L}, \defect{D}, \defect{I}, and \defect{V} combined.  Black lines are the sum over all defect types $T$. For clarity the densities are shown on a semi-logarithmic scale while $n_{ST}$ is shown on a linear scale. Included in the analysis are configurations observed at time $\ta$ in trajectories that proceed through the 57-channel. The corresponding analysis for trajectories that pass through the E-channel can be found in \preapp\ref{app:additional} (\prefig\ref{fig:melting_trajs_defect_spatial_corr_sphere_T_268K_303K_comb_e_channel}).}\label{fig:melting_trajs_defect_spatial_corr_sphere_T_268K_303K_comb}
\end{figure}

We have already shown in the previous section that 5+7 defects tend to form within the eventual volume of the liquid nucleus. \Prefig\ref{fig:melting_trajs_defect_center_i_v} assesses the analogous behavior for mobile defects. While at $T = \SI{268}{\kelvin}$ the first mobile defect that leads to melting forms inside the volume of the future liquid bubble $80\%$ percent of the time, at $T = \SI{303}{\kelvin}$ this share has declined to $64\%$. In both cases, the formation site is correlated with the center of the liquid nucleus (i.e. $g(r)$ is not flat), however, the mobile defects are formed outside the nucleus volume in a significant number of trajectories.

To investigate the role of \defect{5+7} defects in the creation of mobile defects,  \prefig\ref{fig:melting_trajs_defect_spatial_corr_sphere_T_268K_303K_comb} shows an analysis of the defect densities around the site where a mobile defect forms. Only trajectories that proceed via the 57-channel are included in this analysis. Note, that due to the sampling frequency of $(\SI{10}{\pico\second})^{-1}$ \defect{5+7} defects with a life time smaller than \SI{10}{\pico\second} may not be detected in this analysis.

At both temperatures we find that the density of \defect{5+7} defects is enhanced close to the site where a mobile defect forms and we find the closest \defect{5+7} defect within \SI{6}{\angstrom} in $71\%$ of trajectories at \SI{268}{\kelvin} (\SI{303}{\kelvin}: $57\%$). The average number of defects within \SI{6}{\angstrom} is $0.87$ (\SI{303}{\kelvin}: $0.70$). While at $T = \SI{303}{\kelvin}$ all \defect{5+7} defect types roughly contribute equally to the nearby defect population, at $T = \SI{268}{\kelvin}$ there is a preference for \defect{5+7-5} defects.

In trajectories that pass through the E-channel, mobile defects can also form close to the \defect{E} defect. Hence, due to volume exclusion, the density of \defect{5+7} defects around the site where a mobile defect forms is suppressed relative to the densities found in trajecotories that pass through the 57-channel (see \preapp\ref{app:additional}, \prefig\ref{fig:melting_trajs_defect_spatial_corr_sphere_T_268K_303K_comb_e_channel}).

\begin{figure}[tbp]
  \begin{center}
    \includegraphics[width=1.0\columnwidth]{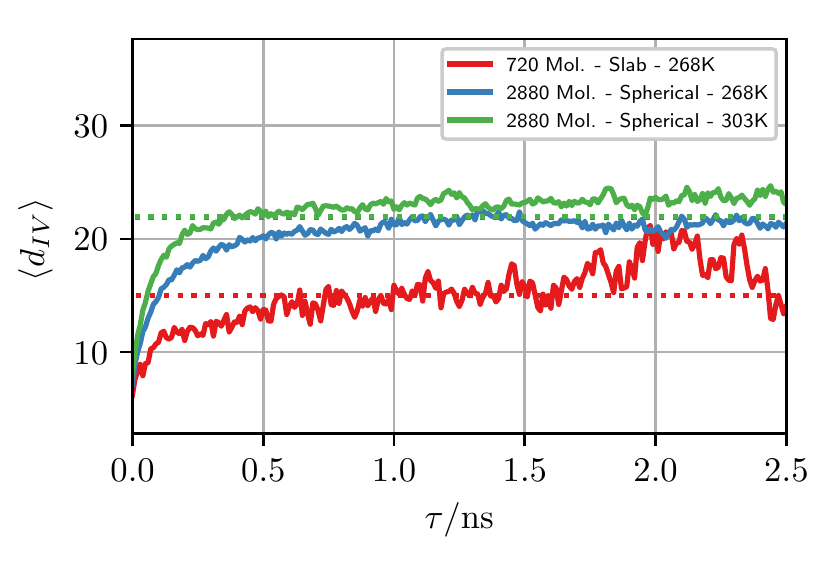}
  \end{center}
  \caption{Average distance between interstitial and vacancy as a function of time after their first creation in melting trajectories. Shown are different simulation box- and cluster geometries and temperatures. The data shown has been calculated considering configurations with a single \defect{I-V} pair because the assignment of defects into pairs is unambiguous in this case. The dashed horizontal lines indicate the average distance between two randomly chosen points in the simulation box of the respective geometry.}\label{fig:i_v_distances}
\end{figure}

\begin{figure}[tbp]
  \begin{center}
    \includegraphics[width=1.0\columnwidth]{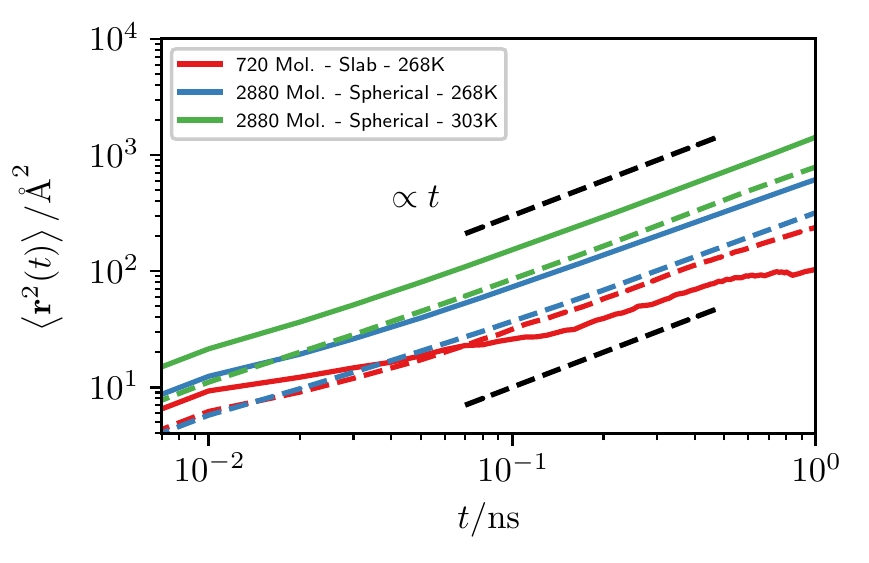}
  \end{center}
  \caption{Mean squared displacement of interstitial (solid lines) and vacancy defects (dashed lines) as observed in between the time of their first formation in melting trajectories up to the time where an extended liquid cluster forms, $\tm$. Different colors represent different system sizes and temperatures, and the dashed black lines are a guide to the eye indicating linear behavior. Refer to \prefig\ref{fig:i_1d_MSDs_slab_only} in the appendix for a breakdown of $\left<\vec{r}^2(t)\right>$ into different directional components $\left<r_i^2(t)\right>$.}\label{fig:i_v_msds}
\end{figure}

After an \defect{I-V} defect pair has formed we can track the motion of the two defects through the system. In \prefig\ref{fig:i_v_distances} we show the average distances between the \defect{I} and the \defect{V} defects as a function of the elapsed time since the \defect{I-V} pair has formed. The average is calculated over all configurations where a single I-V pair is present and each trajectory is included until time $\tm$. Within roughly \SI{0.5}{\nano\second} the average distance between \defect{I} and \defect{V} defects approaches the value expected if one were to randomly place two particles in the simulation box (dotted lines in \prefig\ref{fig:i_v_distances}). In \prefig\ref{fig:i_v_msds} we show the mean-square-displacements, $\left< \vec{r}^2(t) \right>$, of \defect{I} and \defect{V} defects during the same timespan. After a subdiffusive regime that lasts roughly \SI{100}{\pico\second}, $\left< \vec{r}^2 (t) \right>$ is close to linear indicating that the defects freely diffuse through the system. At \SI{268}{\kelvin} the ratio of the self-diffusion constants of intersitials and vacancies, $D_I / D_V$, equals approximately $2$ (\SI{303}{\kelvin}: $1.75$).

Together these two datasets suggest that in the timespan between $\te$ and $\tm$ both defect types independently diffuse through the simulation box. Indeed we find that the non-equilibrium pair correlation function between interstitials and vacancies obtained from configurations observed in this timespan is flat for distances larger than \SI{10}{\angstrom} (see \preapp\ref{app:additional}, \prefig\ref{fig:iv_dist_distributions}).

\begin{figure}[tbp]
  \begin{center}
    \includegraphics[width=1.0\columnwidth]{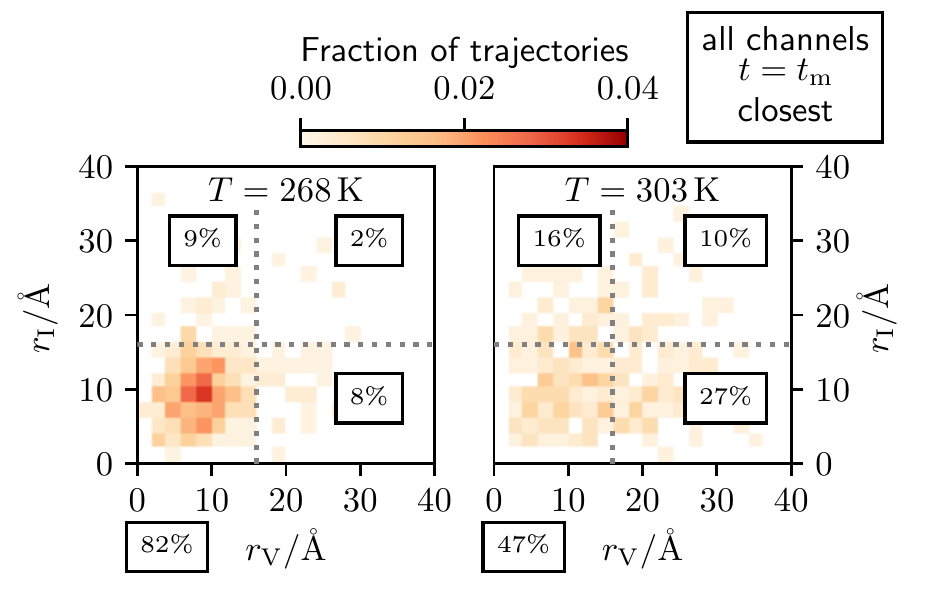}
  \end{center}
  \caption{Analysis of the locations of \defect{I-V} defects at the time of formation of a liquid nucleus, $\tm$, using the center of the nucleus volume as reference. Shown are the fractions of trajectories in which the closest vacancy and the closest interstitial are found at distances $r_\mathrm{V}$ and $r_\mathrm{I}$,respectively. The gray dotted lines indicate distances of \SI{16}{\angstrom}, just larger than the radius of the liquid nucleus volume of \SI{15}{\angstrom}. The percentages indicate the fraction of trajectories where the distances fall within the areas outlined by gray dotted lines. The percentage of trajectories where both defects are within \SI{16}{\angstrom} of the center of the bubble is given in the lower left corner.} \label{fig:melting_trajs_defect_center_i_v_tm}
\end{figure}

To further investigate the role of \defect{I} and \defect{V} defects in the formation of a liquid nucleus, \prefig\ref{fig:melting_trajs_defect_center_i_v_tm} shows the position of the \defect{I-V} defect pair that is closest to the center of the nucleus volume at time $\tm$. We find that at \SI{268}{\kelvin} in $98\%$ of trajectories there is at least a single \defect{I} or \defect{V} defect present inside the volume that later becomes the liquid nucleus. In $82\%$ of trajectories both the closest \defect{I} and \defect{V} defect are found in this volume. No significant imbalance between the two types of defects can be detected.

Note that \prefig\ref{fig:melting_trajs_defect_center_i_v_tm} shows the probabilities of finding the \defect{I} and \defect{V} defect closest to the center of the liquid nucleus. Even for an ideal gas, the analogous distribution of the closest gas atom to a given location is not flat but has a maximum at a distance that is determined by the density of the gas. In \preapp\ref{app:closest_particle_ideal_gas} we calculate the expected shape of this distribution and compare it to the data shown in \prefig\ref{fig:melting_trajs_defect_center_i_v_tm}. We find that there is a strong excess of \defect{I} and \defect{V} defects that are close to the liquid nucleus' center at time $\tm$ relative to the expected distances based on uncorrelated density fluctuations.

At \SI{303}{\kelvin} the distribution of defect positions at time $\tm$ is broader. Here we find that in $90\%$ of trajectories at least one \defect{I} or \defect{V} defect is close to the center of the liquid nucleus at time $\tm$. \defect{V} defects are found outside of the nucleus volume slightly more often than \defect{I} defects. Note, that the change in slope of $n_5(t)$ at time $\tm$ is less pronounced at $T = \SI{303}{\kelvin}$ than at $\SI{268}{\kelvin}$, making the determination of $\tm$ less precise. The broader distribution of defect positions is at least in part a consequence of this reduced precision. Nevertheless, also here we find a strong excess of \defect{I} and \defect{V} defects close to the center of the nucleus volume compared to a uniform distribution of defect positions.

\subsection{Analysis of trajectories with slab shaped clusters}\label{sec:differences_slab}
\begin{figure}[tbp]
  \begin{center}
    \includegraphics[width=1.0\columnwidth]{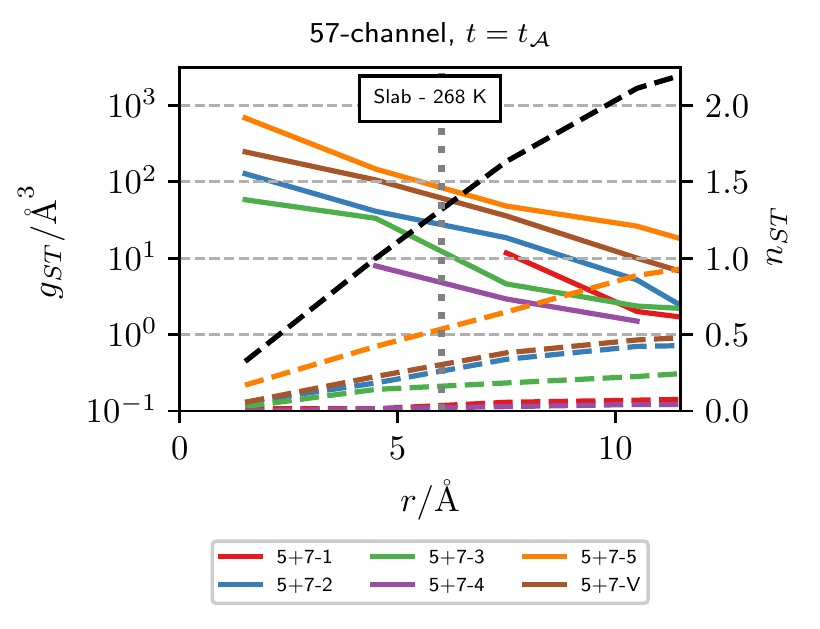}
  \end{center}
  \caption{Same analysis as shown in \prefig\ref{fig:melting_trajs_defect_spatial_corr_sphere_T_268K_303K_comb} for trajectories that end in a slab-shaped cluster in a 720 molecule system. }\label{fig:melting_trajs_defect_spatial_corr_slab_T_268K}
\end{figure}

The results obtained for slab-geometry trajectories at $T = \SI{268}{\kelvin}$ are largely similar to the ones obtained for spherical clusters at the same temperature. The transition proceeds via the E-channel in $57 \pm 7 \%$ of trajectories, via the 57-channel in $42 \pm 6 \%$ and via the D-channel in $1 \pm 1 \%$ of trajectories. This is in good agreement with the data obtained using spherical nuclei. Furthermore, also in this case we find that mobile defects are preferentially formed around \defect{5+7} defects and in particular around \defect{5+7-5} defects, as can be seen in \prefig\ref{fig:melting_trajs_defect_spatial_corr_slab_T_268K}. \Prefig\ref{fig:defect_stats_aligned_on_last_s} demonstrates that the dynamics of forming \defect{I-V} pairs after $\te$ are essentially identical to the ones observed with spherical cluster geometries.

A notable difference can be observed in the mobility of interstitial defects (\prefig\ref{fig:i_v_msds}): while all \defect{I} and \defect{V} defects exhibit some subdiffusivity at timescales below \SI{100}{\pico\second} regardless of cluster geometry and temperature, this effect is strongly enhanced in interstitials that form when the liquid slab decays. The same is not true for the corresponding vacancies. An analysis that separates the contributions to the MSD in different lattice directions (see \preapp\ref{app:additional}, \prefig\ref{fig:i_1d_MSDs_slab_only}) also suggests that the movement of interstitials is hindered in both the direction orthogonal to the secondary prism plane and the direction orthogonal to the basal plane. An analysis of the positions of interstitial defects in these trajectories shows that they are confined to the volume that later becomes the slab shaped liquid domain. This suggests that interstitials are more likely to not leave the volume of the liquid domain between the time they are formed and $\tm$ if the cluster is slab shaped.

\begin{figure}[tbp]
  \begin{center}
    \includegraphics[width=1.0\columnwidth]{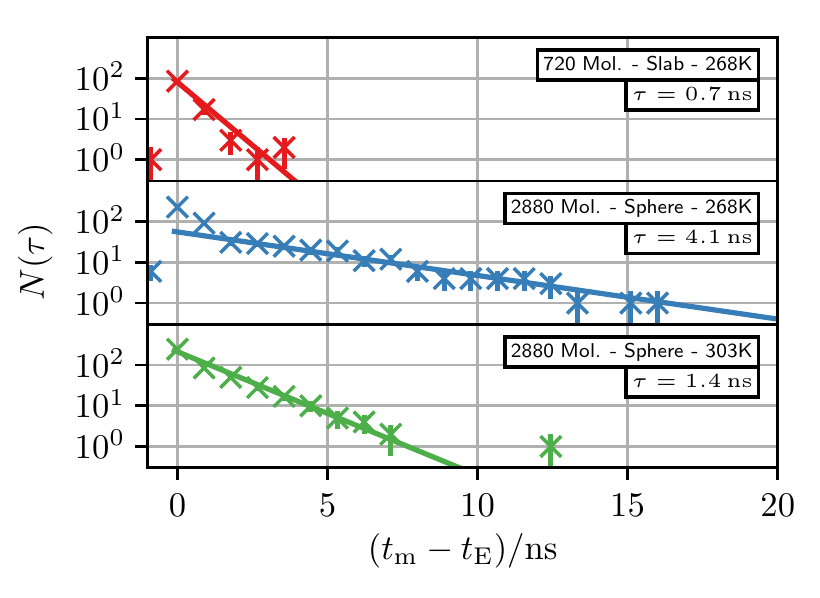}
  \end{center}
  \caption{Distributions of waiting time $\tm - \te$ between the last time melting trajectories contain only defects of type \defect{5+7}, \defect{455778}, or extended defects of type \defect{E} and the formation of an extended liquid cluster. Shown are fits of the normalized exponential distribution $\tau^{-1} e^{-t/\tau}$ to all data points with $\tm - \te > 0$ except for the simulations with a spherical cluster at $T = \SI{268}{\kelvin}$ where only the data points with $\tm - \te > \SI{1.5}{\nano\second}$ are used.}\label{fig:waiting_time_histos}
\end{figure}

We also analyzed the distribution of waiting times between $\te$ and $\tm$ (\prefig\ref{fig:waiting_time_histos}). Here we find a strong dependence of the waiting time on the simulation geometry where (at the same temperature of \SI{268}{\kelvin}) the decay time $\tau$ decreases from \SI{4.1}{\nano\second} in the sphere-geometry simulations to \SI{0.7}{\nano\second} in the slab-geometry. This reduction in waiting time may in part be a consequence of the pinning of interstitials in the slab volume. However, based on calculations by \textcite{LeVot2020} for one-dimensional systems, we would expect a significant dependence of waiting times on system size even in the case of equal diffusivities.

\section{Discussion and outlook}\label{sec:discussion}
We analyzed melting trajectories that were obtained using molecular dynamics simulations at coexistence and at $11\%$ superheating. At superheating conditions we observed the following sequence of events: (i) on average 1.5 \defect{5+7} defects form within the volume that later becomes the liquid nucleus; 0.7 of these defects are \defect{5+7-5} defects; (ii) in roughly 1/3 of trajectories a larger topological defect structure that fulfills the ice rules (an \defect{E} defect) forms; (iii) a mobile defect (an L-D or an I-V pair) forms close to either an \defect{E} if one exists or close to a \defect{5+7} defect; (iv) if an \defect{L-D} pair has formed, an \defect{I-V} pair forms within a timescale of \SI{0.5}{\nano\second}; (v) the \defect{I-V} defects freely diffuse through the system; (vi) a liquid nucleus forms as a result of the interaction of the mobile defects with the \defect{5+7} defects that are already present in the volume that is later occupied by the critical liquid nucleus.

Previous studies\precite\cite{Donadio2005,Mochizuki2013} have pointed out the role of \defect{5+7} defects and larger defect structures in the melting mechanism of ice (modelled by the TIP4P water model). Our results show that the role of \defect{5+7} defects is two-fold: \defect{5+7} defects are often found close to the site where the first mobile defect is formed and, secondly, \defect{5+7} and \defect{E} defects create a defective region that is succeptible to the formation of a liquid nucleus when a mobile interacts with it. 

As the temperature is reduced this defective region tends to become larger as demonstrated by the increase of melting trajectories that pass through the E-channel ($52\%$ at \SI{268}{\kelvin} compared to $35\%$ at \SI{303}{\kelvin}) and the increase in the average number of \defect{5+7} defects found close to the site where the liquid nucleus forms ($2.2$ compared to $1.5$). This finding is consistent with the results of of \textcite{Mochizuki2013} who reported that at $19\%$ superheating \defect{5+7} defects play a role in the formation of mobile defects but no accumulation of these defects occurs prior to melting. 

In \preonlineref\onlinecite{Mochizuki2013} it was also shown that the formation of a separated interstitial-vacancy pair is the rate limiting step in the limit of high superheating. Under the conditions investigated in this paper, the rate limiting step is the formation of a liquid nucleus of critical size, however, also in this case interstitials and vacancies play a crucial role in that they cooperate with \defect{5+7} defects to form the initial liquid nucleus.

Prior to the formation of the liquid nucleus the interstitials and vacancies diffuse freely through the simulation box. This has interesting implications when one wants to extrapolate simulations results to the thermodynamic limit. Since interstitials and vacancies are only weakly bound to each other they are also not limited to form close to the accumulation of immobile defects they later interact with to form a liquid nucleus. This is underscored by the finding that even in the strongly confined environment of our simulation boxes we find that a considerable number ($36 \%$ at \SI{303}{\kelvin}) of mobile defects forms outside of the eventual volume of the liquid nucleus. This suggests that mobile defects may diffuse for considerable distances before they encounter an accumulation of other defects and form a liquid nucleus.

In summary, the results of our simulations further support the observation that interstitials and vacancies play an integral role in the microscopic mechanism of ice melting and establish that, as the degree of superheating becomes smaller, increasingly large immobile defect structures are present within the volume where the initial liquid nucleus forms prior to its formation. Interstitial and vacancy defects then interact with these immobile defects to form an initial liquid nucleus that later grows and melts the ice crystal. This adds ice to the list of solids with a melting mechanism that involves the prior accumulation of defects.

\begin{acknowledgments}
  C.M. has been supported by an uni:docs fellowship of the University of Vienna. C.M. and C.D. acknowledge support from the Austrian Science Fund (FWF) Project No. I3163-N36. P.G. acknowledges the generous support (2/17 to 5/17) of the Erwin Schrödinger Institute for Mathematics and Physics (ESI). P.G. was supported (6/17 to 5/21) by the U.S. Department of Energy, Office of Basic Energy Sciences, through the Chemical Sciences Division (CSD) of Lawrence Berkeley National Laboratory (LBNL), under Contract DE-AC02-05CH11231. The computational results presented have been achieved in part using the Vienna Scientific Cluster (VSC).
\end{acknowledgments}

\section*{Data Availability}
The data that support the findings of this study are available from the corresponding author upon reasonable request.

\section*{References}
\bibliography{paper_slab_relaxation}

\appendix

\section{The defect detection scheme}\label{app:defect_detection_details}
\begin{figure}[tbp]
  \begin{center}
    \includegraphics[width=0.9\columnwidth]{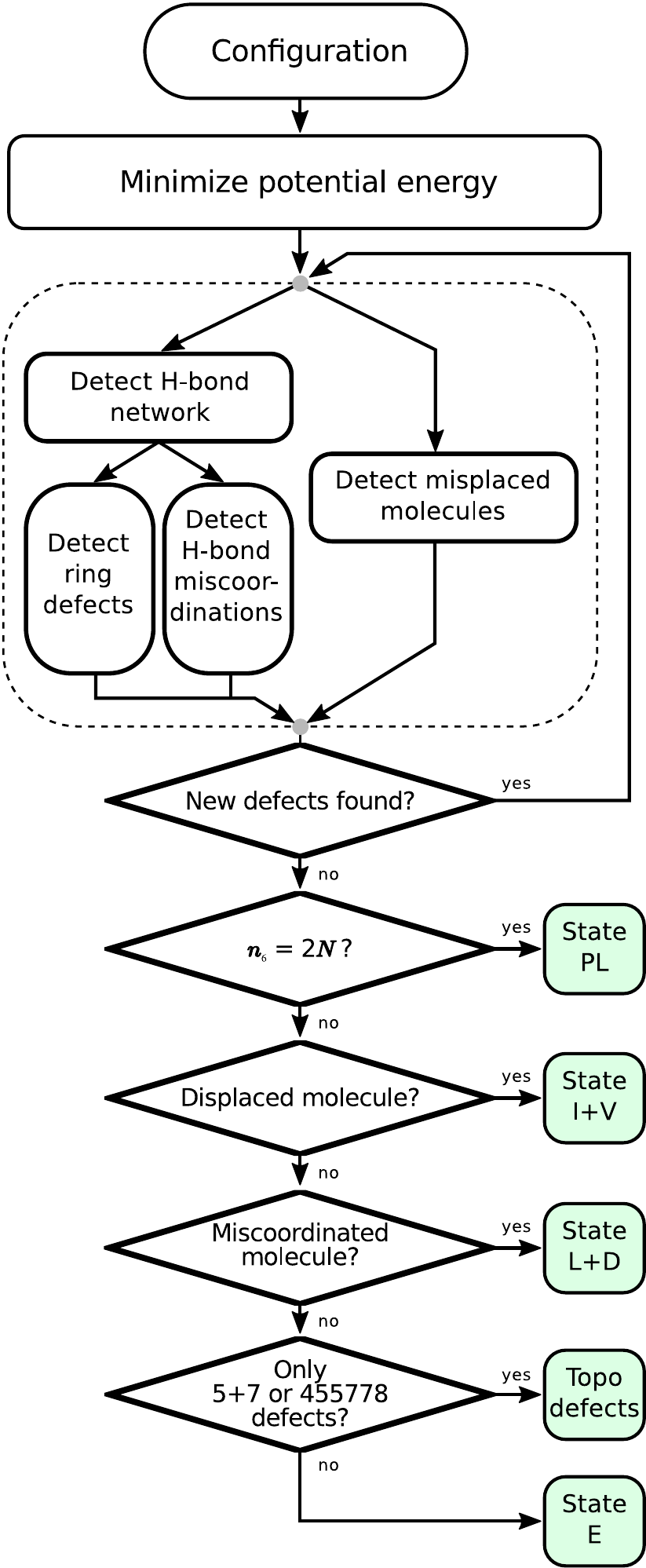}
  \end{center}
  \caption{Sketch of the algorithm used to detect defects in the ice Ih structure throughout this paper. It consists of two stages: in the first stage the configuration is analyzed and deviations from the pristine Ice Ih structure are marked using different approaches (see the descriptions in this appendix). In the second stage a state is assigned based on the deviations found. $n_6$ and $N$ are the number of 6-rings detected and the number of molecules in the system, respectively.}\label{fig:defect_detection_flow_chart}
\end{figure}

\Prefig\ref{fig:defect_detection_flow_chart} presents the scheme used to detect defects in configurations throughout the paper. We start by minimizing the potential energy of a given configuration by setting the momenta of atoms to zero and annealing the system  to a temperature of \SI{1}{\kelvin} using a Langevin thermostat. The cooled configurations are then analyzed using a number of algorithms the details of which can be found below.

In the course of this analysis deviations from a perfect hexagonal ice lattice are detected and, if possible, attributed to known defect structures. The attribution of anomalies in the H-bond network to a defect may reveal other known defect structures and, hence, the analysis is run iteratively until no new defects are found.

The last part of the detection scheme assigns each configuration to a global state based on the criteria shown in \prefig\ref{fig:defect_detection_flow_chart}.

\subsection{Detection of the hydrogen bond network and of L-D defects}
Hydrogen bonds are detected in annealed configurations using the HB2 criterion proposed in \preonlineref\onlinecite{VilaVerde2012} using a custom implementation for the LAMMPS simulation package. This criterion considers a set of two oxygen atoms and one hydrogen atom to form a hydrogen bond if the O-O distance is smaller than \SI{3.5}{\angstrom} and the O-H-O angle is larger than \SI{140}{\degree}. This analysis yields a network of HBs from which an undirected graph is constructed using the \emph{networkx} python package\precite\cite{Aric2008}. This package provides a large number of graph analysis functions we use to carry out parts of the following analysis.

The simplest task is the count of the number of HBs each molecule donates ($n_\text{d}$) and accepts ($n_\text{a}$). In the perfect lattice one expects $n_\text{d} = n_\text{a} = 2$ for all molecules. If either of these two criteria is not fulfilled for a given molecule it is considered to be \emph{miscoordinated}. This can be the result of both, an interstitial-vacancy and an L-D pair. If no \defect{I-V} pair is found, an L-D pair is considered to be present. Note, that \defect{L-D} pairs where \defect{L} and \defect{D} stay bound to each other form frequently in hexagonal ice and that their separation is associated with an additional free energy barrier\precite\cite{Grishina2004}. Here we do not distinguish between bound and unbound \defect{L-D} pairs.

\subsection{Detection of topological defects in the hydrogen bond network}\label{app:ring_analysis}
Defects in the ice Ih structure such as the 5+7 defect can present very similar molecular environments to what can be seen in the perfect lattice. The hydrogen bonds in the region of 5+7 defects fulfill the ice rules and the changes in the bond angles are comparatively small. This makes it hard to detect defects based on the positions of molecules around a given molecule alone. In order to detect such defects, algorithms that are based on detecting the topology of the hydrogen-bond network have been proposed\precite\cite{Donadio2005,Mochizuki2013}. Here we use a similar scheme that analyzes the rings that can be found in the network formed by H-bonds in a hexagonal Ice crystal.

We define a ring in the H-bond network as a path that leads from a molecule to itself, where we move from one molecule to the next following H-bonds. We do not allow the path to cross itself, i.e.\ each intermediate molecule may only be visited once. In principle, with the use of periodic boundary conditions, a given molecule may be part of an infinite number of rings of arbitrary size. To make our ring definition unique we restrict ourselves to a smaller set of rings: the set of shortest rings around each molecule that contain the molecule and two of its neighbors.

\begin{figure}[tbp]
  \begin{center}
    \includegraphics[scale=1.0]{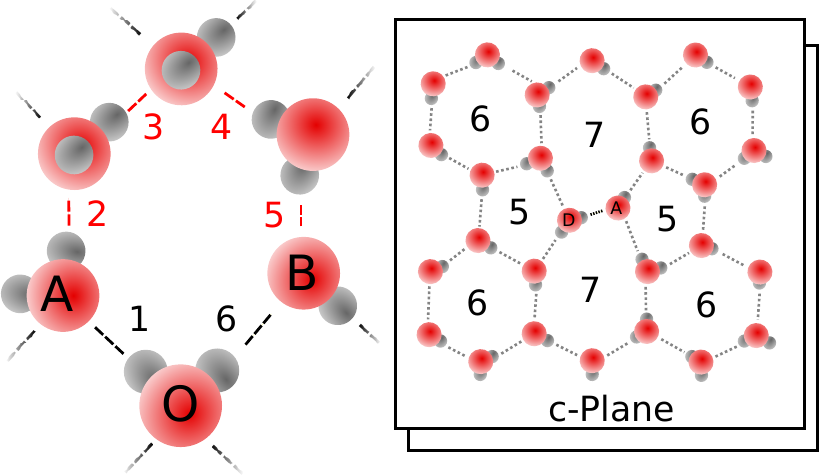}
  \end{center}
  \caption{Illustration of the ring-detection scheme used to detect topological defects in Ice Ih crystal lattice. \figleft rings are detected by selecting a molecule $O$ and all pairs of molecules it is conencted to via hydrogen-bonds, $A$ and $B$. Rings are then found by finding all shortest paths from $A$ to $B$ that do not pass through edges $O-A$ and $O-B$ (H-bonds 1 and 6 respectively). Subsequently the ring is closed by adding molecule O to the ring. \figright Illustration of a 5+7 Type 3 defect\precite\cite{Podeszwa1999a,Grishina2004} as seen projected onto the $c$-plane of the crystal structure. The numbers within the rings count the number of molecules that are within the respective ring. $D$ and $A$ mark what we call the donor- and the acceptor molecule of the 5+7 defect.}\label{fig:ice_hbonds_sketch}
\end{figure}

Consider a molecule $O$ that is H-bonded to $n_\text{hb}$ other molecules as depicted in \prefig\ref{fig:ice_hbonds_sketch}. In order to construct the shortest rings we iterate through all $n_\text{hb}(n_\text{hb}-1)/2$ pairs of neighbor molecules $A$ and $B$. For each of these pairs we look for the shortest path between $A$ and $B$ that do not pass through the molecule $O$. These paths are constructed using Dijsktra's algorithm\cite{Dijkstra1959} as implemented in networkx. There may be more than one path of the same length and---in order to make the result reproducible---we include all rings in the following analysis. Repeating this procedure for all molecules in the crystal yields a set of rings and a list of rings that pass through a given molecule which we use to identify defect types.

5+7 defects\precite\cite{Podeszwa1999a,Grishina2004} can be formed both in the $c$-plane (``horizontally'') or parallel to the prism-, or the secondary-prism plane. The basic ideas for detecting these defects is demonstrated in the right part of \prefig\ref{fig:ice_hbonds_sketch}: the donor and the acceptor molecule of the 5+7 defect must be part of at least two 7-rings and one 5-ring in the $c$-plane alone. If we also consider the rings that pass through molecules in the adjacent plane, donor and acceptor molecules of a horizontal 5+7 defect are part of 8 7-rings, 4 6-rings and 2 5-rings. This ring fingerprint is used to pick out 5+7 defects from a given configuration.
\begin{table}[tb]
    \centering
    \caption{Table of horizontal 5+7 defect types. The 5+7 defects can be distinguished by the orientation of the donor and acceptor molecule. They can either be in-plane (IP, both outgoing hydrogen bonds are formed with molecules in the same $c$-plane) or out-of-plane (OP, one H-bond to a molecule in the same $c$-plane, one with a molecule in an adjacent plane)\cite{Grishina2004}. To distinguish Type 3 from Type 5 defects the orientation of the in-plane $O-H$ vector of both molecules has to be taken into account.}\label{tab:5_7_types}
    \begin{tabular}{|c|c|c|c|} 
        \hline 
        Type &	Donor & Acceptor &	in-plane $O-H$ vectors\\
        \hline 
        1    &   OP    &   OP     &	-\\
        \hline 
        2    &   IP    &   IP     &	-\\
        \hline 
        3    &   IP    &   OP     &	antiparallel\\
        \hline 
        4    &   OP    &   IP     &	-\\
        \hline 
        5    &   IP    &   OP     &	orthogonal\\
        \hline 
    \end{tabular}
\end{table}

In order to further distinguish the different types of 5+7 defects described in Ref.\,\onlinecite{Grishina2004} we also take into account where the other H-bonds of donor and acceptor are pointing (cf. \pretab\ref{tab:5_7_types}). The HBs of each molecule can either both point towards a molecule in the same layer (in-plane, IP) or one of the HBs points towards a molecule in the same plane and the other one to a molecule in an adjacent plane (out-of-plane, OP). Hence, to detect these different types one needs to associate each molecule with a layer of the configuration. Type 3 and Type 5 defects are additionally distinguished by the direction the in-plane H-bonds of donor and acceptor are pointing to.
\begin{figure}[tbp]
  \begin{center}
    \includegraphics[width=0.4\textwidth]{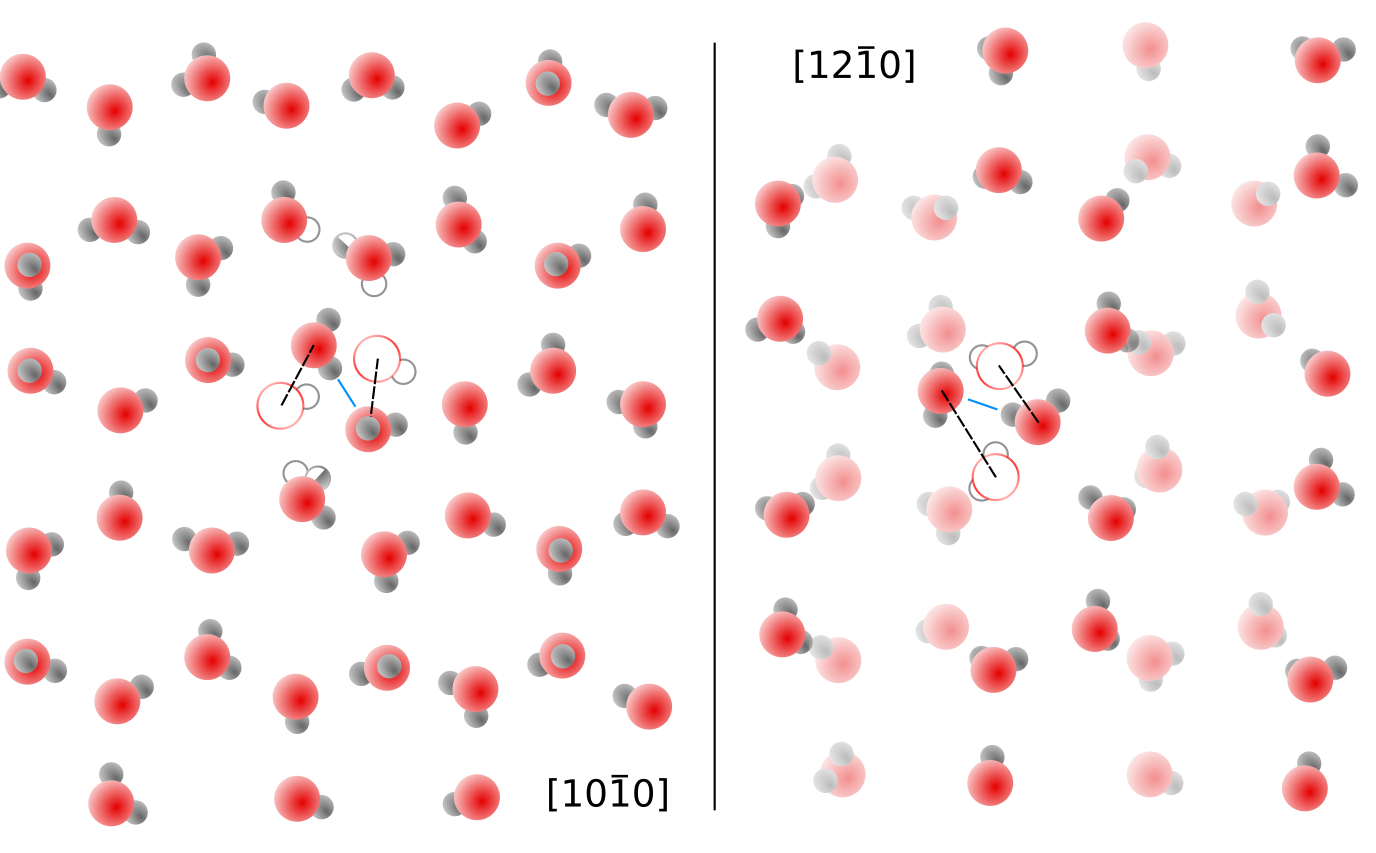}
  \end{center}
  \caption{Sketches of vertical 5+7 defects seen projected onto the prism- ([$12\bar{1}0$]) and the secondary prism plane. The molecules that are only outlined indicate the position of the respective molecule in the perfect lattice. The shaded molecules on the right are roughly \SI{2}{\angstrom} closer to the reader than the unshaded molecules.}\label{fig:57_v_defects_sketch}
\end{figure}

In addition to horizontal 5+7 defects with donor-acceptor pairs that are in the same $c$-plane, also vertical 5+7 defects can be observed. \Prefig\ref{fig:57_v_defects_sketch} shows two examples of such defects where the donor-acceptor pair lies in the prism- ([$10\bar{1}0$]) and the secondary prism-plane ([$12\bar{1}0$]). Together these defects are referred to as \defect{5+7-V} defects.

\begin{figure}[tbp]
  \begin{center}
    \includegraphics[width=0.4\textwidth]{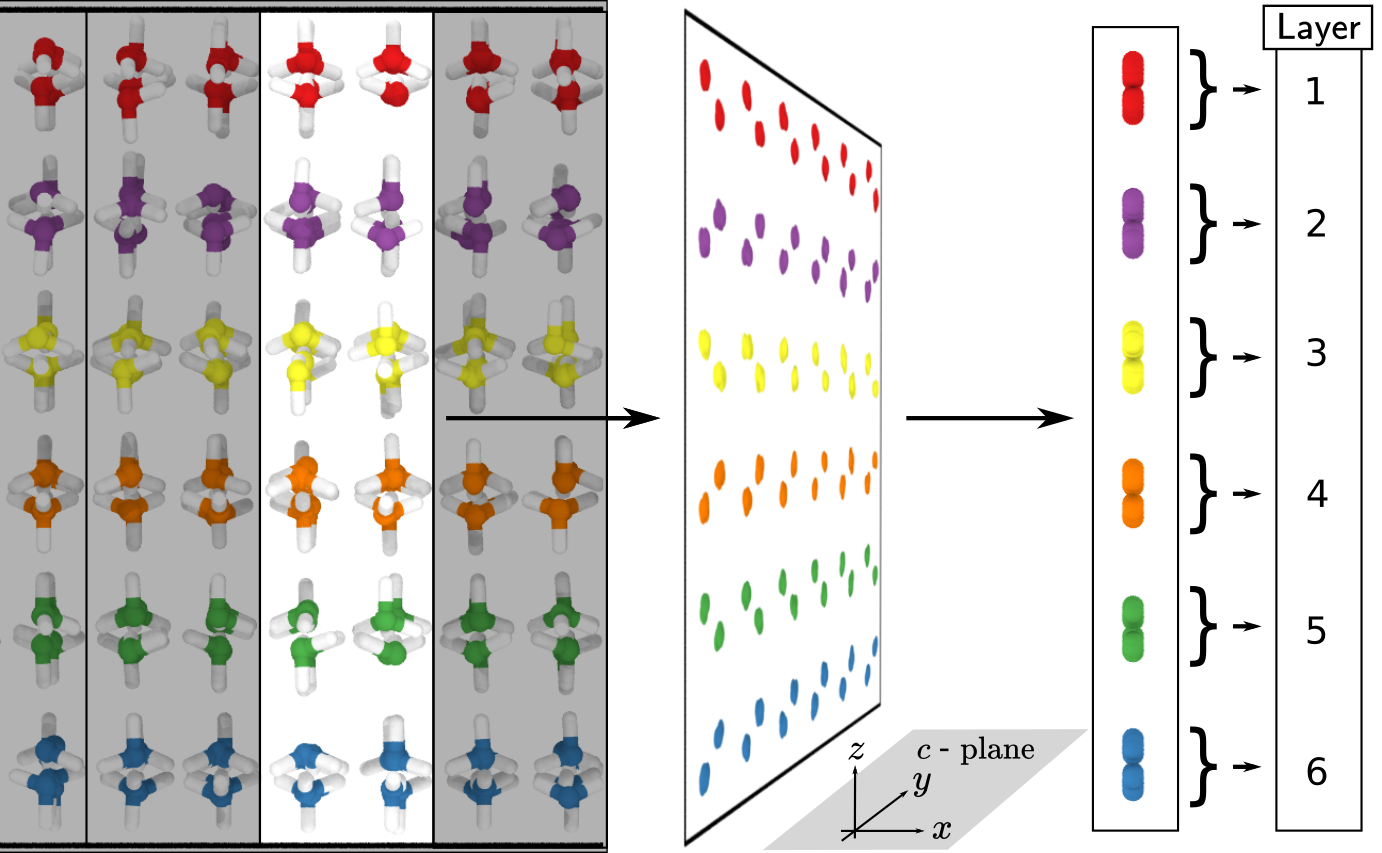}
  \end{center}
  \caption{Schematic representation of the algorithm used to assign molecules to layers. The white area on the left indicates a single slice used for the analysis, which is carried out on each slice separately. The resulting layers are then matched across neighboring slices in order to arrive at a consistent numbering of layers throughout the system.}\label{fig:ice_layer_detection}
\end{figure}

The algorithm that assigns each molecule to a layer is sketched in \prefig\ref{fig:ice_layer_detection}. First, the configuration is sliced along the $x$-axis into slices that contain two layers of molecules. Next, the configuration is squashed leaving only the $z$-coordinates of the oxygen atom of each molecule. These $z$-positions are clustered using the \emph{AgglomerativeClustering} hierarchical clustering algorithm implemented in the scikit-learn python package\precite\cite{scikit-learn}\footnote{scikit-learn version 0.18.2 is used.}. As metric we use the distances between the molecules taking periodic boundary conditions into account. \emph{Average} linkage is used and we set the number of clusters to be found to the number of $c$-planes expected (6 in the example shown). This procedure is applied to each slice along the $x$-direction separately and afterwards the resulting layers are matched across the slices using the center-of-mass of the layers found in each slice to identify which neighboring layers belong together. This slicing procedure is used to accommodate capillary waves that modulate the layer positions as one moves along the $x$-direction. The result of this procedure is that each molecule is now associated with a number that identifies the layer the molecule is in.

The detection of defect structures is complicated if multiple defects are close to each other. In this case, the ring counts of different defects can influence each other. To be able to detect \defect{5+7} defects that are close to each other, we compiled a list of different ring fingerprints by inspecting a large number of configurations with unknown defect structures.

\subsection{Detection of interstiatial-vacancy defects}
In order to identify interstitials and vacancies in configurations we compare the minimized configuration with a template configuration that contains a perfect ice Ih crystal of the same size. This procedure is a simplified version of the method previously employed by \textcite{Mochizuki2013} where we skip the calculation of the edit distance.

\begin{figure}[tbp]
  \begin{center}
    \includegraphics[width=0.4\textwidth]{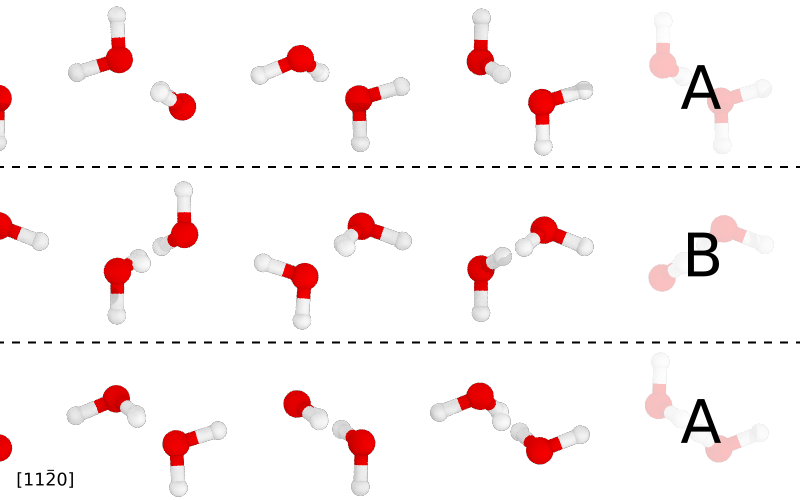}
  \end{center}
  \caption{Ice Ih structure as seen facing the prism face. The structure is comprised of alternating layers A and B with respect to the position of the oxygen atoms. Protons are arranged in a random pattern that fulfills the ice rules.}\label{fig:ice_layers_sketch}
\end{figure}

To do so, we first align the configuration that is analyzed (denoted by $\mathcal{C}$) with the template configuration ($\mathcal{T}$) using the following procedure:
\begin{enumerate}
	\item Identify the set of molecules in the system, $\mathcal{I}$, whose contribution to the potential energy in the system is lower than $\SI{-16}{\kilo\calorie\per\mol}$. These molecules are in ice-like configurations.
	\item Align these molecules with their counterparts in the template configuration. To do so, use a global shift of the configuration, $\vec{s}$, to minimize the summed mean square displacement
\begin{equation}
 	F(\vec{s};\mathcal{C},\mathcal{T}) = \sum_{i \in \mathcal{I}} \left| \vec{x}_i^\mathcal{C}  - \vec{X}(\vec{x}_i^\mathcal{C};\mathcal{T}+\vec{s}) \right|^2,
\end{equation}
wherein $\vec{X}(\vec{x}_i^\mathcal{C};\mathcal{T}+\vec{s})$ is the position of the closest neighbor of molecule $i$ in $\mathcal{C}$ that can be found in $\mathcal{T}$ that has been shifted by $\vec{s}$. As the position of a molecule we use the position of the oxygen. This yields an optimized vector $\vec{s}^{(1)}$ and a value of the MSD function $F_\text{min}^{(1)}$.
\item If $F_\text{min}^{(1)} / N_\text{mol}$ is larger than a threshold value $f_\text{max} = \SI{0.03}{\angstrom\squared}$, shift the vector $\vec{s}^{(1)}$ by one layer distance in the $z$-direction and redo the optimization using this shifted vector as initial condition.
\item If $F_\text{min}^{(2)}/ N_\text{mol}$ is still larger than $f_\text{max}$, again, shift the output vector of the previous configuration and redo the optimization. If this optimization does not succeed, the procedure fails.
\end{enumerate}
\begin{figure*}[p]
  \begin{center}
    \includegraphics[width=0.9\textwidth]{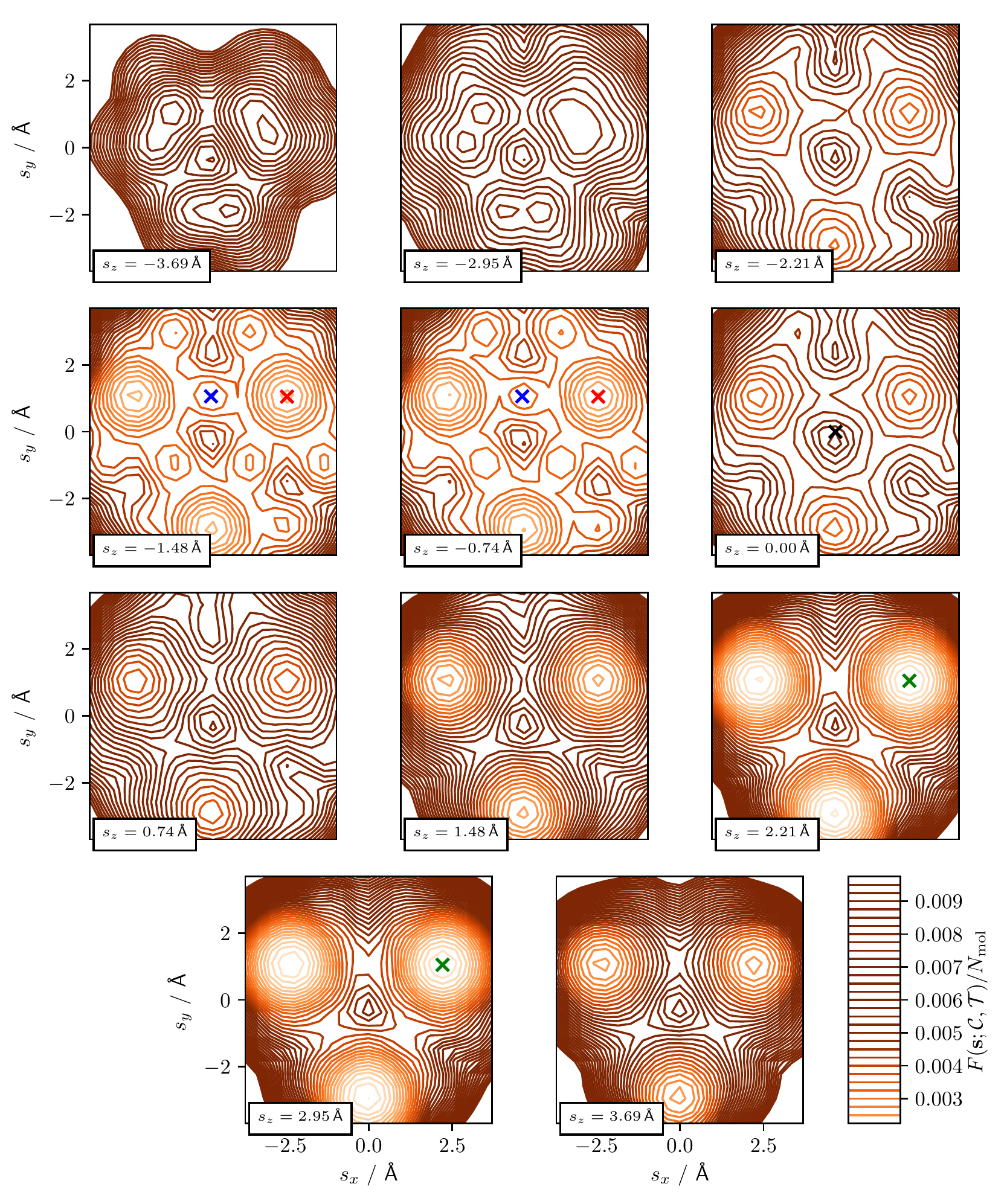}
  \end{center}
  \caption{Example contour plot of summed mean-square deviation $F(\vec{s};\mathcal{C},\mathcal{T})$ as a function of the shift vector $\vec{s}$. The example shows the landscape used to align a configuration that containes $720$ molecules that are arranged in a mostly pristine lattice except for a small number of defects. The black cross indicates the vector $\vec{s}$ at which the first optimization is started. The blue cross indicates the value of $\vec{s}$ after the first optimization ($s^{(1)}_z = -1.14$, in between the two levels shown), which corresponds to a shift of the template configuration relative to the optimal alignment along both the prism- and the c-axis. The red cross indicates another local minimum, which corresponds to an alignment where the crystal is shifted by exactly one layer in the $z$-direction (i.e.\ layers of type A are on top of type B layers). The green cross marks the value of $\vec{s}$ after the second optimization which finds the best alignment of the two configurations ($s^{(2)}_z = 2.56$).}\label{fig:msd_landscape}
\end{figure*}
The reason for this multi-step procedure becomes clear when we examine the landscape $F(\vec{s};\mathcal{C},\mathcal{T})$ shown in \prefig\ref{fig:msd_landscape} that consists of multiple local and global minima. \Prefig\ref{fig:ice_layers_sketch} shows the alternating layers that an hexagonal ice crystal is comprised of. Depending on the initial shift of the template, an optimization of $F$ can lead into the local minimums marked by green and red dots in \prefig\ref{fig:msd_landscape}. These minima correspond to shifts where the template configuration is offset by exactly one layer width along the $z$-direction, i.e.,\ such that the B-layers of the template is found on top of the A-layers of the configuration.

Fortunately, at the densities observed in the simulations presented here, there are no local minima if $\vec{s}$ is chosen such that the layers match. Hence, by performing multiple optimizations where $\vec{s}$ is shifted by one layer in the $z$-direction after the first optimisation, we find the correct alignment of the two configurations. Only in some cases the third optimization is required because the initial $\vec{s}$ of the first optimization finds a small local minimum. The optimizations are performed using the BFGS algorithm as implemented in the scipy python package\cite{Jonesa}.

Given the optimal alignment of $\mathcal{T}$ to $\mathcal{C}$ we can now assign each molecule in $\mathcal{C}$ to the site in $\mathcal{T}$ that is its nearest neighbor and count the number of times that each site in $\mathcal{T}$ is found as a nearest site. In a defect free crystal this count, $n^\mathcal{T}_i$, is equal to one for each site. However, if there are interstitial-vacancy pairs in the system, some of the $n^\mathcal{T}_i$s are different from one. These sites are marked as template mismatches (TMM).

To avoid false positives that come about due to molecules translating only locally (e.g. to form a \defect{5+7} defect), an additional step is performed where, if a site with an excess molecule and a site with a missing molecule are neighbors of each other (in the sense that they are H-bonded to each other in the template configuration), this pair of template mismatches is ignored.

The TMM that remain are marked as interstitials if $n^\mathcal{T}_i > 1$ and as vacancies if $n^\mathcal{T}_i < 1$.

\section{Ideal gas closest distance distribution}\label{app:closest_particle_ideal_gas}
In \prefig\ref{fig:melting_trajs_defect_center_i_v_tm} (\presec\ref{sec:results}) we presented the histograms of the distances of the closest interstitial and the closest vacancy to the center of the liquid domain. It is important to note that even for a homogeneous distribution of defects (i.e. an ideal gas), the distribution of the closest defect is not uniform but instead has a maximum that is determined by the density of the gas and the volume of the simulation box\precite\cite{Hertz1909}.

Consider the probability of finding the atom closest to a given location at a distance that falls into the interval $\left[r, r+\text{d}r\right]$ for an ideal gas. Because the atom positions are not correlated with each other in the ideal gas, this probability can be decomposed into three factors: the probability that a specific atom is found at a distance inside $\left[r, r+\text{d}r\right]$ times the probability that none of the $N-1$ other atoms are closer than $r + \text{d} r$, times the number of particles that can be picked. With the total system volume $V$ and the sphere volume $v(r) = 4\pi r^3 / 3$ the probability density $p_\text{id}$ can be written as
\begin{equation}\label{equ:ideal_gas_joint_density}
  \begin{aligned}
    p_\text{id}(r) \, \text{d}r &= \left(\frac{V - v(r)}{V}\right)^{N-1} \frac{4 \pi r^2}{V} \, N \, \text{d}r \\
                      &= 4 \pi \rho r^2 \exp\left[(N-1) \log\left(1 - v(r)/V\right)\right] \text{d}r.
  \end{aligned}
\end{equation}
In the thermodynamic limit ($N \to \infty$, $V \to \infty$ such that $N/V = \rho = \text{const.}$) this can be approximated by
\begin{equation}
  p_\text{id}(r) \, \text{d}r \approx 4 \pi \rho r^2 \exp\left[-4 \pi \rho r^3/3\right].
\end{equation}

To calculate the joint probability that the closest ideal gas-like interstitial is found at a distance in the interval $[r_\text{I},r_\text{I} + \text{d}r_\text{I}]$ and the closest ideal-gas vacancy in the interval $[r_\text{V},r_\text{V} + \text{d}r_\text{V}]$ we multiply the two probabilities:
\begin{equation}
  p_\text{id}(r_\text{V}, r_\text{I}) \, \text{d}r_\text{V} \, \text{d}r_\text{I} = p_\text{V}(r_\text{V}) p_\text{I}(r_\text{I}) \, \text{d}r_\text{V} \, \text{d}r_\text{I}
\end{equation}

Integration over the size of a histogram bin, $\Delta r$, yields the probabilities
\begin{equation}
  P(r_\text{V}, r_\text{I}) = \int_{r_\text{V}}^{r_\text{V} +  \Delta r} \int_{r_\text{I}}^{r_\text{I} +  \Delta r} \text{d}r_\text{V} \, \text{d}r_\text{I} \, p_\text{V}(r_\text{V}) p_\text{I}(r_\text{I})
\end{equation}

To compare this expected distribution to the ones obtained from melting trajectories we need to determine the average densities $\rho_\text{I}$ and $\rho_\text{V}$ at time $\tm$. To do so we count the total number of defects of type $i$ observed in configurations at time $\tm$, $n_i$, and divide by the number number of trajectories $n$ and the average volume of the simulation box $V$.

\begin{figure}[tbp]
  \begin{center}
    \includegraphics[width=1.0\columnwidth]{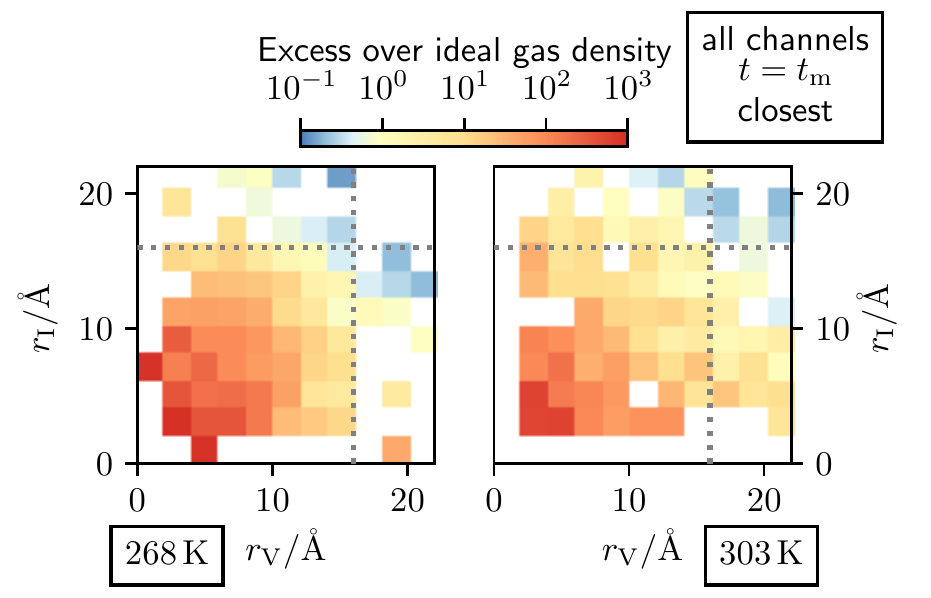}
  \end{center}
  \caption{Analysis of the locations of \defect{I-V} defects at the time of formation of a liquid nucleus, $\tm$, using the center of the nucleus volume as reference. Shown are the ratios between the observed joint probability of the closest vacancy and the closest interstitial, $P(r_\text{V}, r_\text{I})$, to the expected density assuming the defects follow ideal gas statistics $P_\text{id}(r_\text{V}, r_\text{I})$. $P(r_\text{V}, r_\text{I})$ is shown in \prefig\ref{fig:melting_trajs_defect_center_i_v_tm} in the main part.
  } \label{fig:melting_trajs_defect_center_i_v_tm_excess}
\end{figure}

\Prefig\ref{fig:melting_trajs_defect_center_i_v_tm_excess} shows the ratio between the observed probabilities, $P(r_\text{V}, r_\text{I})$, and the ideal gas probabilities $P_\text{id}(r_\text{V}, r_\text{I})$. There is a strong excess of configurations where interstitials and vacancies are close to the center of the liquid nucleus compared to what is expected in an ideal gas.

\onecolumngrid
\section{Data supplement}\label{app:additional}
Here we present a number of additional graphs that were not included in the main part for clarity. 

\begin{itemize}
  \item \Prefig\ref{fig:first_in_A_defect_stats_268K} shows the immobile defect counts found at time $\ta$ in spherical-geometry trajectories obtained at \SI{268}{\kelvin} and \SI{303}{\kelvin}.
\begin{figure*}[tbp]
  \begin{center}
    \includegraphics[width=0.45\textwidth]{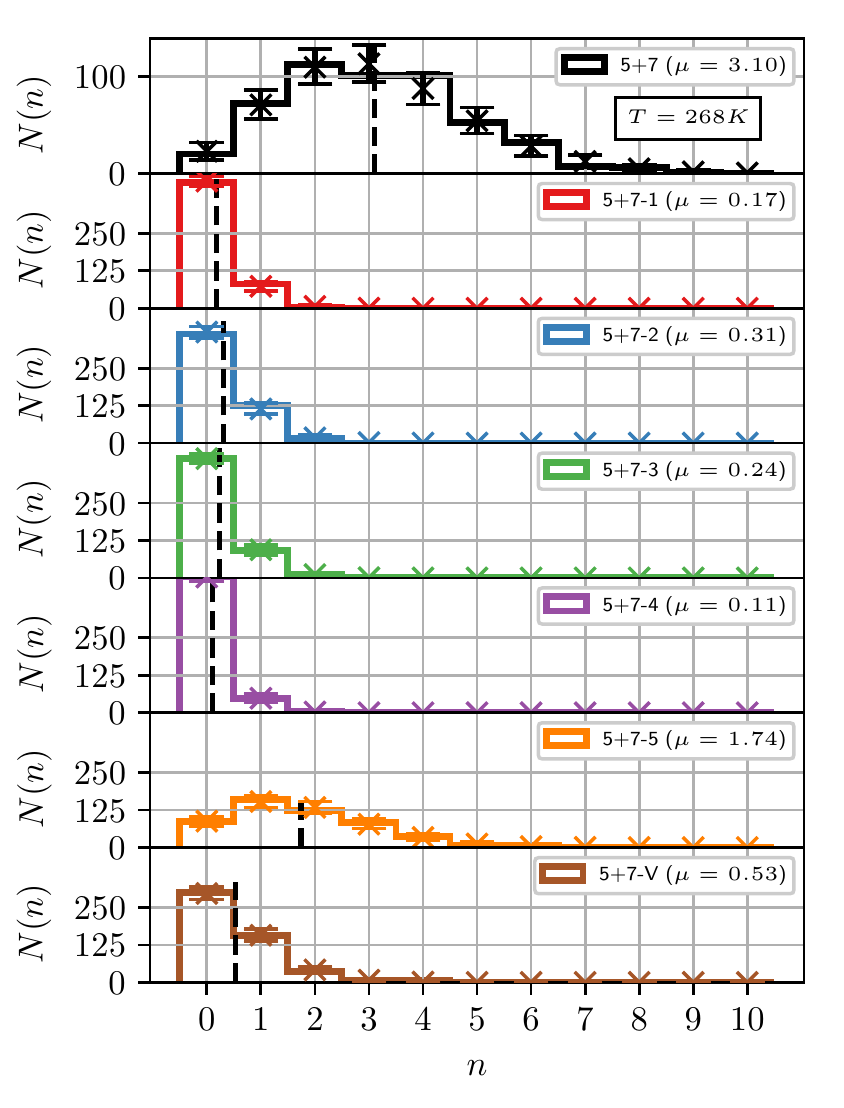}
      \includegraphics[width=0.475\columnwidth]{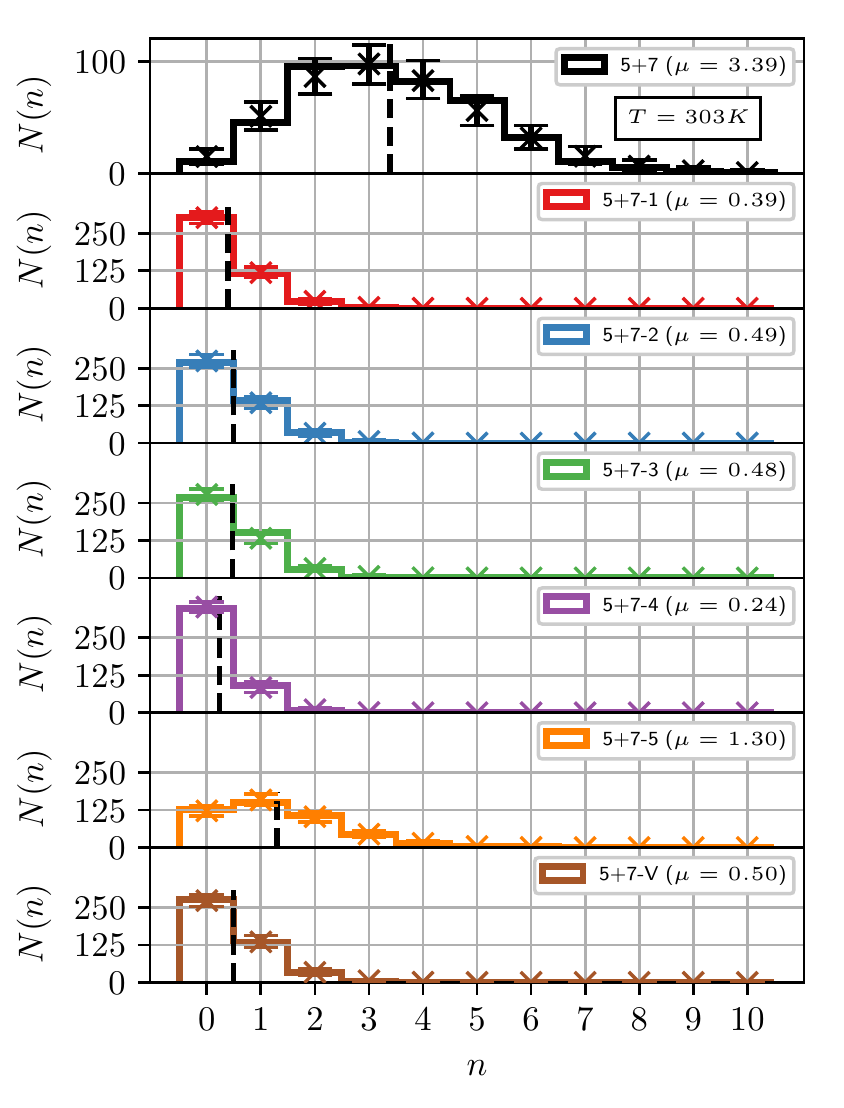}
  \end{center}
  \caption{Statistics of the number of immobile defects of given types found at time $\ta$ in freezing/melting trajectories that end in a spherical nucleuls at temperature $T = \SI{268}{\kelvin}$ (left) and $T = \SI{303}{\kelvin}$ (right). The histograms indicate the counts observed in trajectories and the vertical dashed line indicates the average defect count, $\mu$. The crosses represent the expected counts based on a Poisson distribution with average count $\mu$ and the errorbars indicate the interval into which $95 \%$ of counts would fall based on this distribution.}\label{fig:first_in_A_defect_stats_268K}
\end{figure*}

  \item \Prefig\ref{fig:melting_trajs_defect_spatial_corr_sphere_T_268K_303K_comb_e_channel} shows the defect densities of \defect{5+7} defects around the site where a mobile defect forms for trajecotires that pass through the E-channel. See \prefigs\ref{fig:melting_trajs_defect_spatial_corr_sphere_T_268K_303K_comb} and \ref{fig:melting_trajs_defect_spatial_corr_slab_T_268K} for the corresponding plots that include all trajectories.
\begin{figure}[tbp]
  \begin{center}
    \includegraphics[width=0.475\columnwidth]{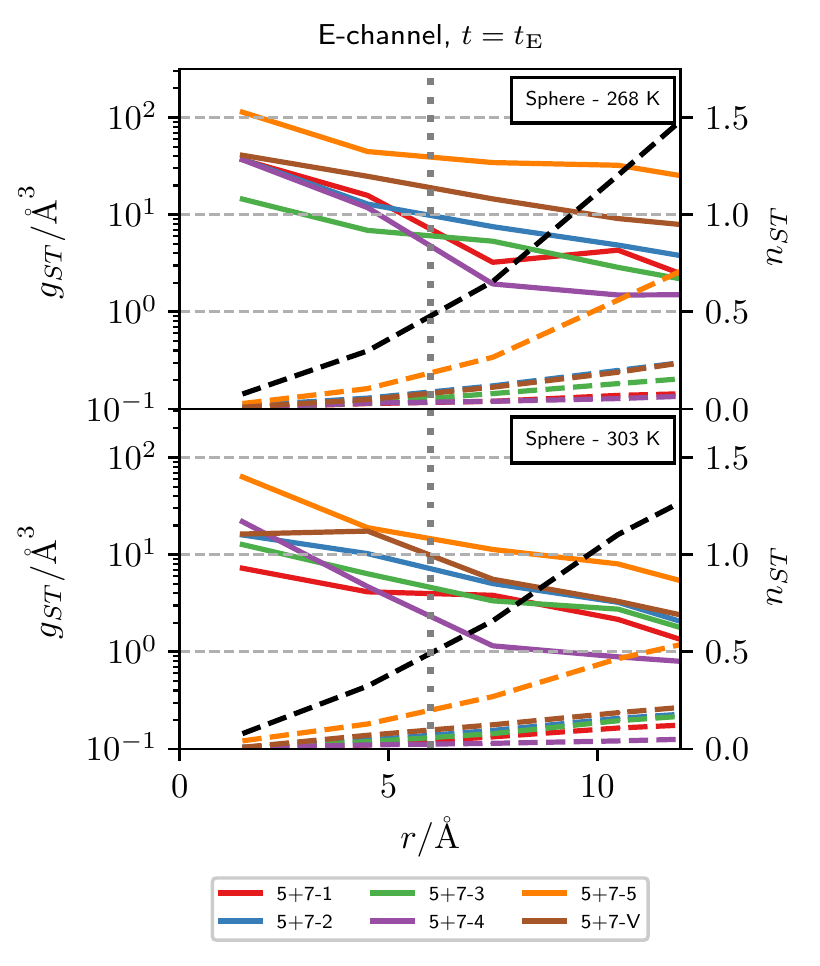}
  \end{center}
  \caption{Analysis of the positions of the closest defect to the site where  \defect{L-D} or \defect{I-V} defects form at time $\te$ in trajectories that pass through the \defect{E}-channel. Shown are the pair-correlation functions $g_{ST}$ (solid) and the average numbers of defects within a sphere of radius $r$, $n_{ST}$ (dashed). Black lines are the sum over all defect types $T$. For clarity the densities are shown on a semi-logarithmic scale while $n_{ST}$ is shown on a linear scale.}\label{fig:melting_trajs_defect_spatial_corr_sphere_T_268K_303K_comb_e_channel}
\end{figure}

  \item \Prefig\ref{fig:iv_dist_distributions} shows the non-equlibrium pair correlation functions for \defect{I} and \defect{V} defects obtained in the timespan between $\te$ and $\tm$.
\begin{figure*}[tbp]
  \begin{center}
    \includegraphics[width=0.95\columnwidth]{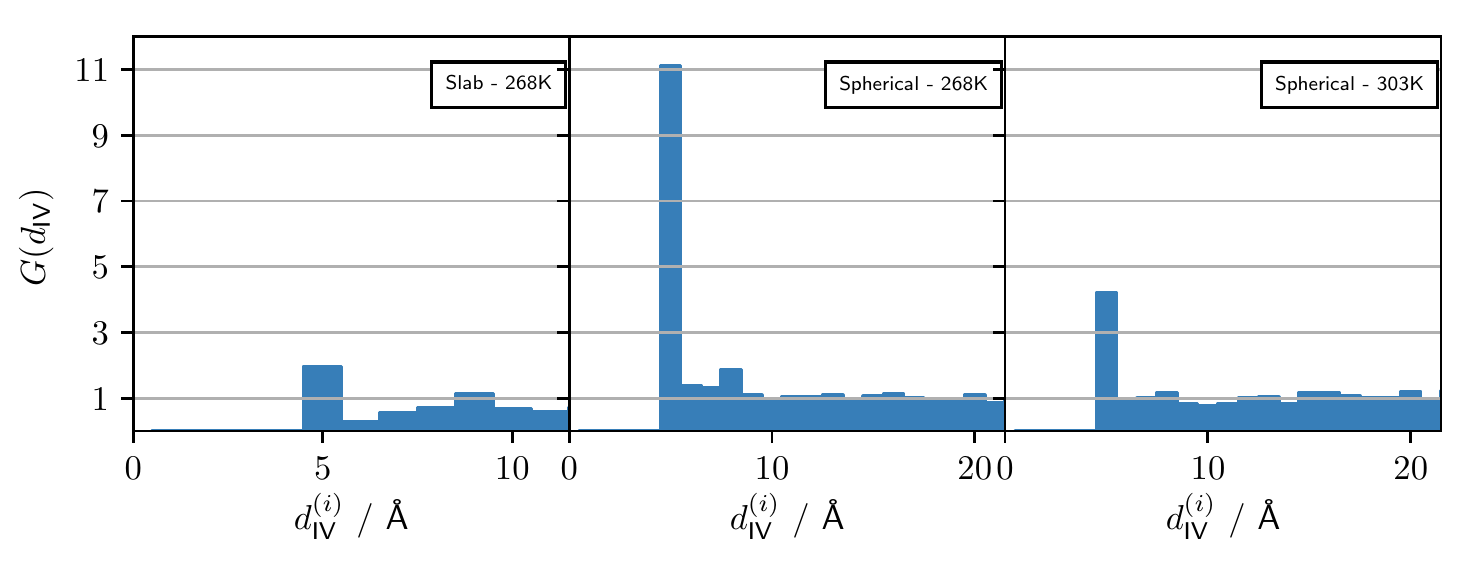}
  \end{center}
  \caption{Pair correlation functions $G$ between interstitial and vacancy defects found in melting trajectories between time $\te$ and $\tm$. Note, that only configurations where a single defect pair is present are included in the analysis and that the trajectories in the underlying ensemble start at the creation of a defect pair. As a consequence, $G$ is not an equilibrium property since finding defect pairs a short distance apart is guaranteed. The distribution has been normalized with the density $\rho = V^{-1}$, where $V$ is the box volume resulting in an assumed number density of the defects of $V^{-1}$.}\label{fig:iv_dist_distributions}
\end{figure*}

  \item \Prefig\ref{fig:i_1d_MSDs_slab_only} shows the one-dimensional contributions to the mean square displacements of \defect{I} and \defect{V} defects in the same time period but only when there is a single \defect{I-V} pair present.
\begin{figure}[p]
  \begin{center}
    \includegraphics[width=0.95\columnwidth]{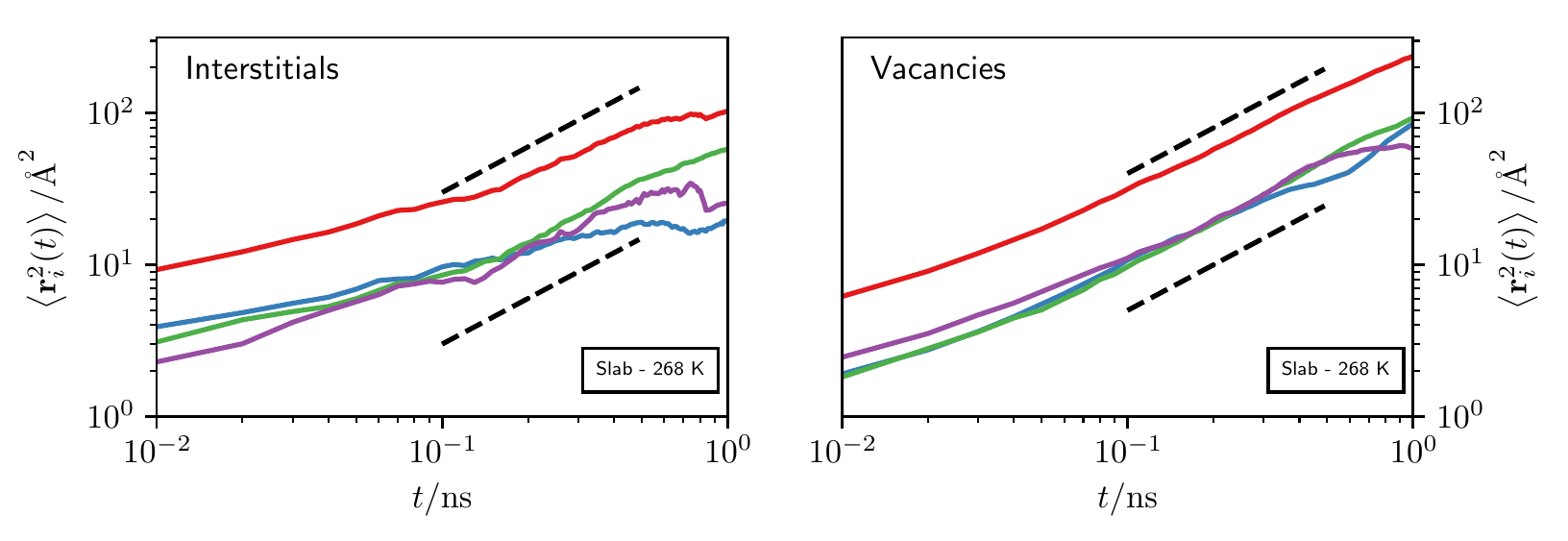}
    \includegraphics[width=0.95\columnwidth]{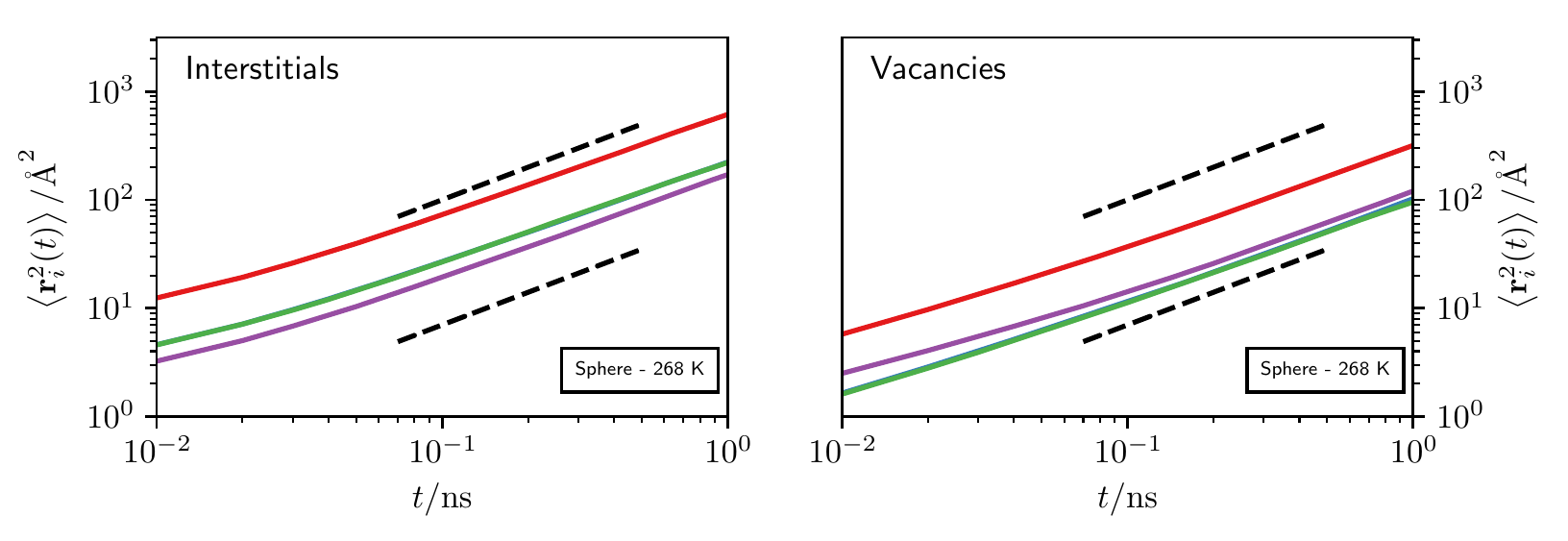}
    \includegraphics[width=0.95\columnwidth]{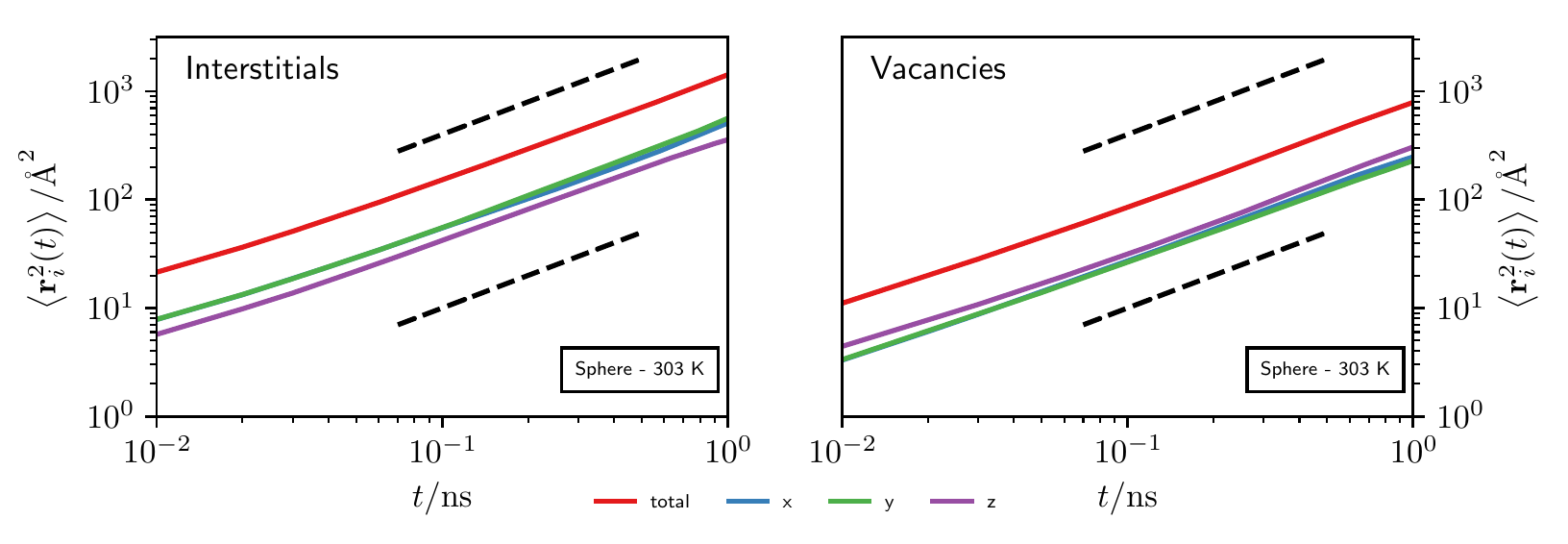}
  \end{center}
  \caption{Mean square displacements calculated for interstitials and vacancies at different temperatures and for different cluster geometries. The $x$, $y$, and $z$ component are shown separately. The dashed lines indicate diffusive behavior where $r^2(t) \sim t$.}\label{fig:i_1d_MSDs_slab_only}
\end{figure}
\end{itemize}

\end{document}